\documentclass[pra,twocolumn,nofootinbib,floatfix,10pt]{revtex4-2}

\usepackage{amsmath}
\usepackage{amssymb}
\usepackage{wasysym}
\usepackage{graphicx}
\usepackage{color,soul}
\usepackage{physics}
\usepackage{siunitx}
\usepackage{dsfont}
\usepackage{float}
\usepackage[english]{babel}
\usepackage{blindtext}
\usepackage[english,nomargin,inline,marginclue,draft]{fixme}
\pdfpageheight\paperheight
\pdfpagewidth\paperwidth

\usepackage[colorlinks,linkcolor=blue,anchorcolor=blue,citecolor=blue,urlcolor=blue]{hyperref}

\fxusetheme{colorsig}
\FXRegisterAuthor{cg}{acg}{CG}  
\FXRegisterAuthor{th}{ath}{\color{blue}TH}  
\FXRegisterAuthor{ib}{aib}{\color{red}IB} 
\FXRegisterAuthor{sh}{ash}{\color{cyan}SH} 
\FXRegisterAuthor{db}{adb}{\color{green}DB} 
\FXRegisterAuthor{ps}{aps}{PS}
\makeatletter
\renewcommand*\FXLayoutInline[3]{%
  {\@fxuseface{inline}\ignorespaces{\color{fx#1}[#3: #2]}}}
\makeatother

\long\def\symbolfootnote[#1]#2{\begingroup%
\def\thefootnote{\fnsymbol{footnote}}\footnotetext[#1]{#2}\endgroup}

\def\nobreakbefore{%
  \relax\ifvmode\else
    \ifhmode
      \ifdim\lastskip > 0pt\relax
        \unskip\nobreakspace
      \else 
        \nobreakspace
      \fi
    \fi
  \fi
}
\let\oldcite\cite
\renewcommand\cite{\nobreakbefore\oldcite}

\begin{document}
\title{Observation of Higgs and Goldstone modes in U(1) symmetry-broken \\ Rydberg atomic systems}

\author{Bang Liu$^{1,2}$}
\author{Li-Hua Zhang$^{1,2}$}
\author{Ya-Jun Wang$^{1,2}$}
\author{Jun Zhang$^{1,2}$}
\author{Qi-Feng Wang$^{1,2}$}
\author{Yu Ma$^{1,2}$}
\author{Tian-Yu Han$^{1,2}$}
\author{Zheng-Yuan Zhang$^{1,2}$}
\author{Shi-Yao Shao$^{1,2}$}
\author{Qing Li$^{1,2}$}
\author{Han-Chao Chen$^{1,2}$}
\author{Jia-Dou Nan$^{1,2}$}
\author{Dong-Yang Zhu$^{1,2}$}
\author{Yi-Ming Yin$^{1,2}$}
\author{Bao-Sen Shi$^{1,2}$}
\author{Dong-Sheng Ding$^{1,2,\textcolor{blue}{\dag}}$}

\affiliation{$^1$Key Laboratory of Quantum Information, University of Science and Technology of China; Hefei, Anhui 230026, China.}
\affiliation{$^2$Synergetic Innovation Center of Quantum Information and Quantum Physics, University of Science and Technology of China; Hefei, Anhui 230026, China.}

\date{\today}

\symbolfootnote[2]{dds@ustc.edu.cn}

\maketitle

\textbf{Higgs and Goldstone modes manifest as fluctuations in the order parameter of system, offering insights into its phase transitions and symmetry properties. Exploring the dynamics of these collective excitations in a Rydberg atoms system advances various branches of condensed matter, particle physics, and cosmology. Here, we report an experimental signature of Higgs and Goldstone modes in a U(1) symmetry-broken Rydberg atomic gases. By constructing two probe fields to excite atoms, we observe the distinct phase and amplitude fluctuations of Rydberg atoms collective excitations under the particle-hole symmetry. Due to the van der Waals interactions between the Rydberg atoms, we detect a symmetric variance spectrum divided by the divergent regime and phase boundary, capturing the full dynamics of the additional Higgs and Goldstone modes. Studying the Higgs and Goldstone modes in Rydberg atoms allows us to explore fundamental aspects of quantum phase transitions and symmetry breaking phenomena, while leveraging the unique properties of these highly interacting systems to uncover new physics and potential applications in quantum simulation.}

The Higgs and Goldstone modes are collective excited modes arising from spontaneous symmetry breaking \cite{goldstone1961field,nambu1961dynamical,goldstone1962broken,anderson1963plasmons,englert1964broken,weinberg1967model,sooryakumar1980raman,littlewood1982amplitude,varma2002higgs, pekker2015amplitude} in condensed matter physics. The Higgs mode involves amplitude fluctuations of a complex order parameter that restore continuous symmetry, which is associated with the massive excitation. The Goldstone mode corresponds to phase fluctuations that manifests as massless excitations that preserve the broken continuous symmetry into a discrete subgroup \cite{strand2010transition}. The Higgs modes refer to excitations associated with the Higgs field, which is a fundamental component of the Standard Model of particle physics. In various contexts, the concept of “Higgs modes" may extend beyond particle physics, sometimes relating to collective excitations in condensed matter systems that exhibit similar characteristics of symmetry breaking and mass generation. The analogy between fundamental particles and many-body systems unveils profound insights into the nature of Higgs modes and symmetry breaking, and plays critical roles in understanding the dynamics of field theories. There are many progresses in observing exotic features of Higgs modes and Goldstone modes in superconductors \cite{matsunaga2014light, sherman2015higgs}, ultra-cold atoms \cite{pollet2012higgs,endres2012higgs,leonard2017monitoring,behrle2018higgs,guo2019low, geier2021exciting}, and antiferromagnets \cite{hong2017higgs}.

The study of Rydberg atomic systems, characterized by their long-range interactions and precisely tuned by external fields \cite{saffman2010quantum,adams2019rydberg,browaeys2020many}, exhibits a versatile platform for investigating non-equilibrium dynamics and emergent phenomena in quantum systems \cite{lee2012collective,carr2013nonequilibrium,schempp2014full,marcuzzi2014universal,lesanovsky2014out, urvoy2015strongly, ding2022enhanced}. The unique properties of Rydberg atoms facilitate the exploration of various exotic phases, including those associated with collective behavior and criticality \cite{ding2022enhanced}, self-organization \cite{Signatures2020Helmrich,ding2019Phase,Wintermantel2020Cellular, klocke2021hydrodynamic}, ergodicity breaking and time crystals \cite{gambetta2019, Wadenpfuhl2023Synchronization, ding2024ergodicity, wu2023,liu2024higher,liu2024bifurcation}. In particular, the emergence of collective modes in thermal interacting Rydberg atoms could provide insights into symmetry breaking mechanisms akin to high energy physics in high-temperature, particularly in scenarios like phase transitions in the early universe, where symmetries can be broken during cooling processes, leading to the emergence of massive particles and other excitations. In this scenario, the long-rang interaction strength changes the effective potential of Rydberg atomic systems, providing a platform to study nontrivial effects in symmetry breaking.

\begin{figure*}
\centering
\includegraphics[width=2.08\columnwidth]{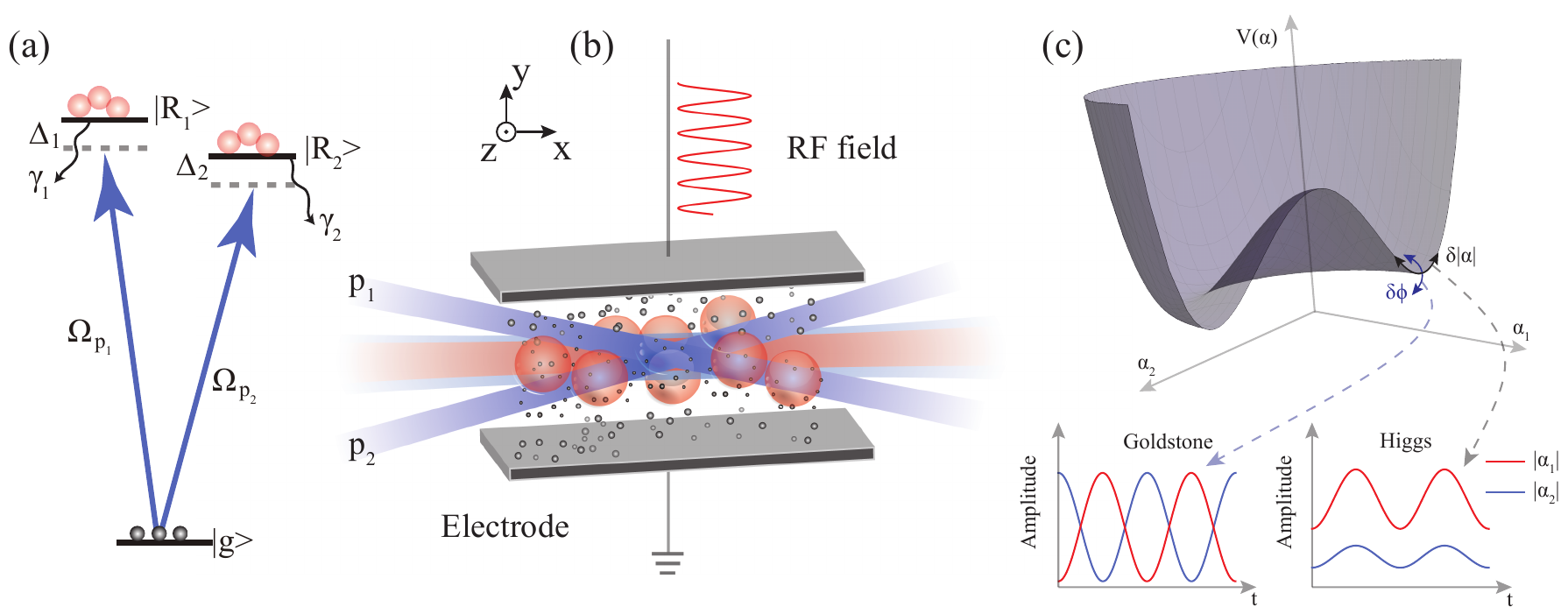}\\
\caption{\textbf{Diagram for observing Higgs and Goldstone modes.} (a) Energy level diagram based on double two-level atomic system. Two lasers with Rabi frequency $\Omega_{p_{1,2}}$ drive the transition from the cesium atom ground state $\ket{g}$ to the Rydberg states $\ket{R_{1,2}}$, respectively. (b) Simplified experimental setup. An RF field is applied to the atoms by two electrodes with a loading frequency of $2\pi\times$7.1 MHz. The external RF electric field strength is $E_{\text{RF}}=U/4$ $V/cm$, $U$ is the applied voltage of two electrodes. (c) The effective potential $V_{\text{eff}}(\alpha) = \xi |\alpha|^2 + \lambda |\alpha|^4$ in ordered phases [with $\xi$ = -10 and $\lambda$ = 10] is described by the complex order parameter $\alpha=|\alpha|e^{i\phi}$. In the phase of symmetry breaking, the fluctuations of amplitude $\delta|\alpha|$ and phase $\delta\phi$ of the order parameter correspond to Higgs and Goldstone modes.}
\label{setup}
\end{figure*}

In this work, we observe the signature characteristics of Higgs and Goldstone modes within the framework of U(1) symmetry-broken Rydberg atomic systems. By employing experimental observations and theoretical models, we aim to highlight the unique features of these modes under long-range interaction and the driving of external external electric field. Our investigation seeks to bridge concepts from particle physics and condensed matter theory in the picture of many-body interaction, offering a comprehensive understanding of how these modes can be manipulated and detected in Rydberg atomic systems. The measured full dynamics of these modes against systematic parameters not only reflect the underlying symmetries of the system but also provide the unique role of many-body interaction in inducing Higgs and Goldstone modes. Through this, we provide more parameter ranges for producing these modes, contributing to broader applications for symmetry breaking of quantum many-body systems.

 \section*{RESULTS}
\subsection*{Physical model}
To generate Higgs and Goldstone modes, we consider two coupled systems with $N_{1(2)}$ interacting double two-level atoms with a ground state \(\ket{g}\) and Rydberg states \(\ket{R_{1,2}}\) (with decay rates \(\gamma_{1,2}\)), see Fig.~\ref{setup}(a). Two probe lasers are used to excite the atoms to Rydberg states in a thermal vapor, with the lasers overlapped at an angle of approximately \(3.6^\circ\) as shown by Fig.~\ref{setup}(b). This is governed by a particle-hole symmetric Hamiltonian, see more details in Methods section. The position-dependent van der Waals interaction, \(V = {C_6}/{r^6}\) (where \(C_6\) is the coefficient and \(r\) represents the distance between Rydberg atoms), leads to distinct phases in the system as the number of Rydberg atoms varies, such as the normal phase and the limit cycle phase \cite{ding2024ergodicity}. The driven and dissipative Rydberg atoms exhibit time translation symmetry breaking, allowing each probe beam to induce criticality and continuous time crystals \cite{wu2023observation,liu2024bifurcation}. The cross-interactions among the excited Rydberg atoms and the competition of the shared ground state atoms in the overlapped volume introduce fluctuations.

The fluctuations do not influence the mean total number of Rydberg atoms within the cross-interaction volume when the interactions between different portions of excited Rydberg atoms are weak. However, when the cross-interaction becomes strong, the total number of Rydberg atoms and the photons numbers \(N =N_1 + 
N_2= \alpha_1^2+\alpha_2^2\) remains nearly constant, and the ratio of excited Rydberg atoms along each probe field direction can vary significantly. This leads to a continuous symmetry breaking of the complex order parameter $\alpha =\alpha_1+i\alpha_2 $ in a Mexican hat-shaped potential with a form of $V_{\text{eff}}(\alpha) = \xi |\alpha|^2 + \lambda |\alpha|^4$ \cite{pekker2015amplitude}, where the real and imaginary parts, $\alpha_1$ and $\alpha_2$ characterize the order parameters associated with each probe field amplitude. The fluctuations of the real and the imaginary parts reveal that there are Higgs and Goldstone modes. By continuously monitoring the order parameter through detection of the probes transmission, we can observe variations in the atoms excitation modulation along the $x$-axis for different field ratios along each direction (Fig.~\ref{setup}(c)).

\begin{figure*}
\centering
\includegraphics[width=2.08\columnwidth]{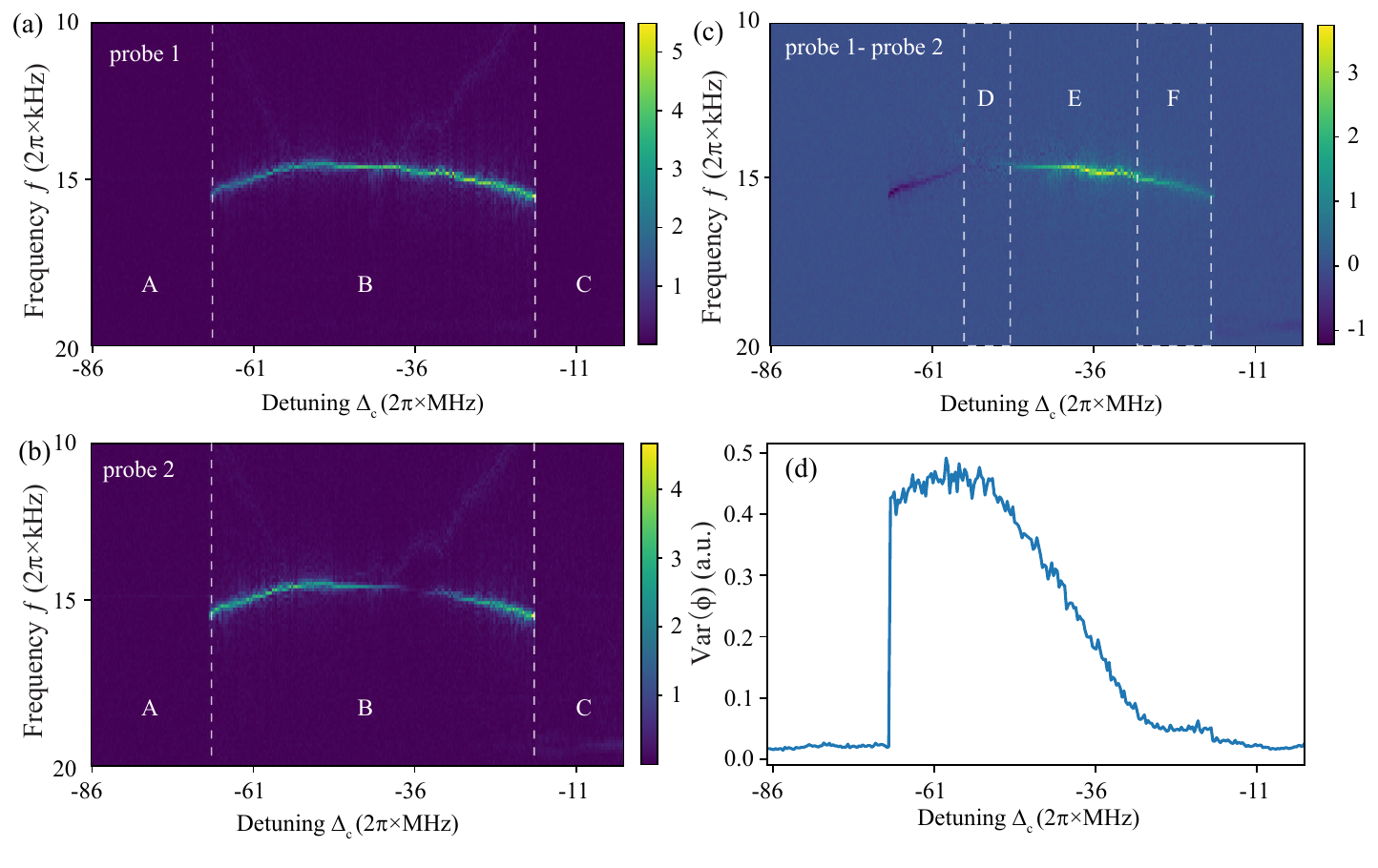}\\
\caption{\textbf{Measured complex phase diagram}. Measured Fourier spectrum of probe 1 (a) and probe 2 (b) versus the coupling detuning $\Delta_{c}$. Labels A-C in panels (a) and (b) indicate regions that define the normal phase (A and C) and symmetry-broken phase (B). (c) shows the Fourier spectrum obtained by subtracting the data in (a) from that in (b), highlighting amplitude fluctuations in relation to the coupling detuning $\Delta_{c}$. Labels D-F in panel (c) denote the regions where the Goldstone and Higgs modes are predominant. Panel (d) illustrates the variance of the phase $\phi$. In these cases, the amplitude of RF-field is $E_{\text{RF}}=0.4$ $V/cm$.}
\label{complex phase diagram}
\end{figure*}

\subsection*{Complex phase diagram}
To show the cross-interactions between the Rydberg collective excitations along probe 1 and probe 2 directions, we map the complex phase diagram of system and extract the conditions of particle-hole symmetry breaking. The complex phase diagram could provide the real and imaginary parts of order parameters, as well as their fluctuations. In experiment, we use a three-photon electromagnetically-induced transparency (EIT) scheme to study the collective modes of Rydberg atoms excitation \cite{zhangRydberg,liuHighly,liu2024bifurcation}. The excitation process involves use of a resonant 852 nm probe beam (a Rabi frequency $\Omega_p$), a resonant 1470 nm dressing beam (with a Rabi frequency $\Omega_d$), and a 780 nm coupling beam (with a Rabi frequency $\Omega_c$ and detuning $\Delta_c$), which drive cesium atoms from the ground state $\ket{6S_{1/2}}$ to the Rydberg state $\ket{49P_{3/2}}$ through the intermediate states $\ket{6P_{3/2}}$ and $\ket{7S_{1/2}}$. 

By varying the coupling detuning $\Delta_{c}$, we record Fourier spectrum of probe transmissions and measure the real part of phase diagram, as shown in Figs.~\ref{complex phase diagram}(a) and (b). Due to the cross-interactions between the Rydberg atoms excited by probe 1 and probe 2, the Rydberg collective excitations become collaboratively synchronized when the system crosses the critical point $\Delta^c_{c} = -2\pi\times 67.5$  MHz, see results in Figs.~\ref{complex phase diagram}(a) and (b). The phase diagrams for probes 1 and probe 2 exhibit identical frequency evolution trajectories and share the same critical point. At these critical points, the system bifurcates into a non-equilibrium state, entering a synchronized time crystal phase within the limit cycle regime \cite{liu2024bifurcation,liuHighly}. In this scenario, the time translation symmetry of the system response is spontaneously broken, in which the system settles into a stable, repeating trajectory in its phase space.

\begin{figure*}
\centering
\includegraphics[width=2.08\columnwidth]{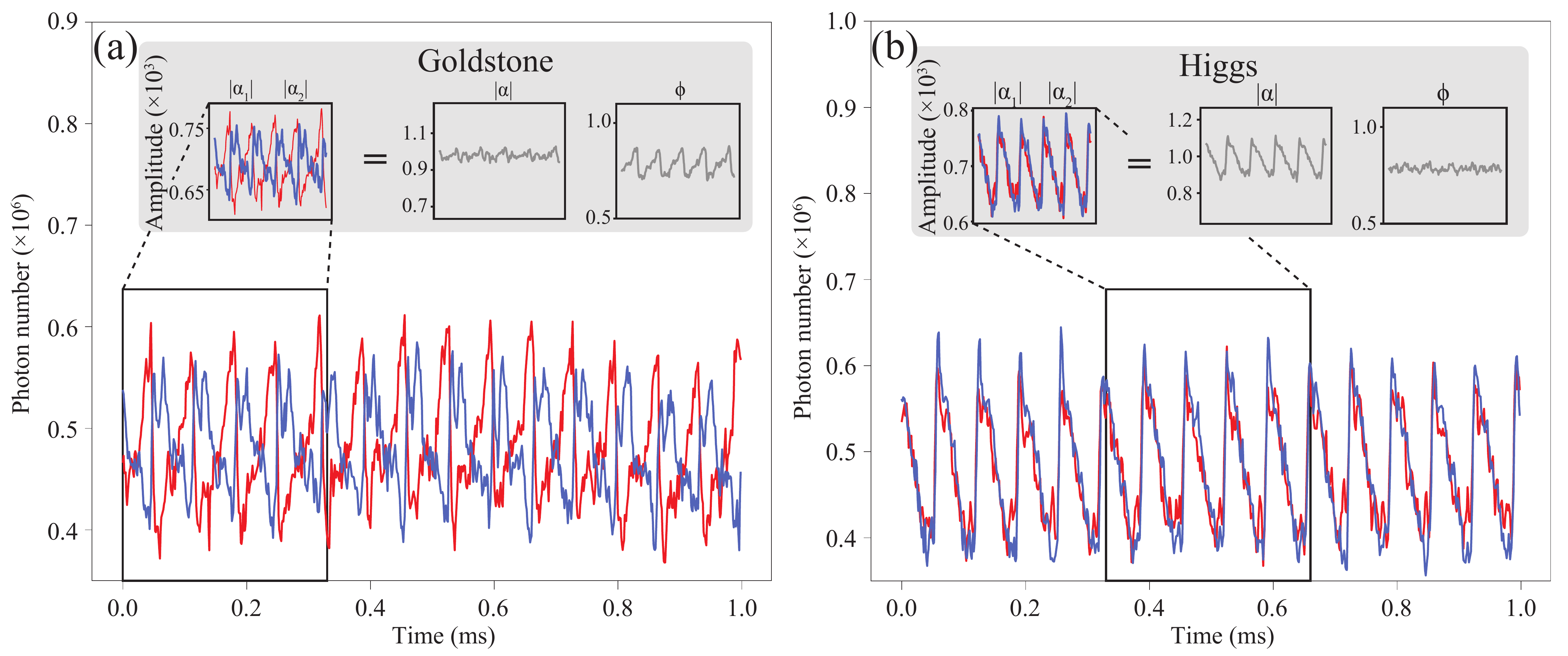}\\
\caption{\textbf{Dynamics of Higgs and Goldstone modes}. The response photon numbers in probe 1 and probe 2 transmission. (a) The measured evolution of Rydberg atom excitation at $\Delta_{c} = -2\pi\times 53.4$ MHz. In polar coordinates, the order parameter displays phase fluctuation $\delta\phi$ with no significant change in amplitude. (b) The measured evolution of Rydberg atom excitation at $\Delta_{c} = -2\pi\times 23.8$  MHz. The inset presents the extracted data over a 0.25 ms interval. By transforming the order parameter into polar coordinates, we observe oscillations in the amplitude $\delta|\alpha|$ while the phase remains nearly constant. }
\label{Higgs and Goldstone}
\end{figure*}

Furthermore, we subtracted the data from the Fourier spectrum of probes 1 and 2 to obtain the fluctuations of amplitude versus the coupling detuning $\Delta_{c}$ [Fig.~\ref{complex phase diagram}(c)]. We find that the significant fluctuations of the Fourier transformed data occur. This reveals the difference of the order parameters, as the amplitude of the data after Fourier transformation is correlated with the amplitude of the transmission oscillation in the time domain. In addition, we measure the imaginary part of the phase diagram to investigate the phase information, see Figs.~\ref{complex phase diagram}(d). We compare the transmission oscillation curves and measure the relative phase difference between them which is defined as $\phi = \text{arctan}(|\alpha_1|/|\alpha_2|)$. Then we calculate the variance of $\phi $, see the results in Fig.~\ref{complex phase diagram}(d). 

When $\Delta_{c} > -2\pi\times 19.5$  MHz or $\Delta_{c} < -2\pi\times 67.5$  MHz, the excitations of Rydberg atoms are found to be in nearly constant for both probes, characterized by small phase fluctuations. However, when $-2\pi\times 19.5$ MHz $<\Delta_{c} < -2\pi\times 67.5$  MHz, the system is in lower time translation symmetry and tends to become ordered in the time domain, leading to an increase in phase fluctuations. By comparing Fig.~\ref{complex phase diagram}(c) and Fig.~\ref{complex phase diagram}(d), we characterize the full dynamics of amplitude and phase fluctuations for the two collective excitation modes, which aids in revealing the dynamics of the Goldstone and Higgs modes.

\subsection*{Higgs and Goldstone modes}
To investigate the Goldstone mode, we tune the system in the D regime in Fig.~\ref{complex phase diagram}(c) by setting $\Delta_{c} = -2\pi\times 53.4$  MHz and record the time flow of probe 1 and probe 2, as shown in Fig.~\ref{Higgs and Goldstone}(a). Transforming the data into polar coordinates reveals that the order parameter experiences phase fluctuations $\delta\phi$, while the amplitude remains relatively stable. In this phase, even though the amplitude of the order parameter may still be non-zero, the phase is no longer fixed, reflecting a loss of long-range order. These phase fluctuations signify the emergence of Goldstone mode—massless excitations [$m_{\text{Goldstone}}=0$] associated with the continuous symmetry.

By varying the coupling detuning to $\Delta_{c} = -2\pi\times 23.8$ MHz, we observe the two probe transmissions in Fig.~\ref{Higgs and Goldstone}(b). In Fig.~\ref{Higgs and Goldstone}(b), the amplitude of the complex order parameter is dynamically oscillated in time domain while the phase remains a nearly constant quantity, resulting in synchronized dynamics. This behavior implies that the system response has transitioned into an ordered phase, where long-range correlations emerge, and the fluctuations become coherent. 

In this ordered state, the response of both probes exhibits time translation symmetry breaking [or lower symmetry], meaning that the system's behavior is not invariant under shifts in time. This is indicative of a collective excitation—specifically, the Higgs mode [a massive excitation with $m_{\text{Higgs}}\neq0$], which corresponds to amplitude fluctuations of the order parameter $\delta |\alpha|$ and also reflects changes in the excited Rydberg atoms number. 

\begin{figure*}
\centering
\includegraphics[width=2.1\columnwidth]{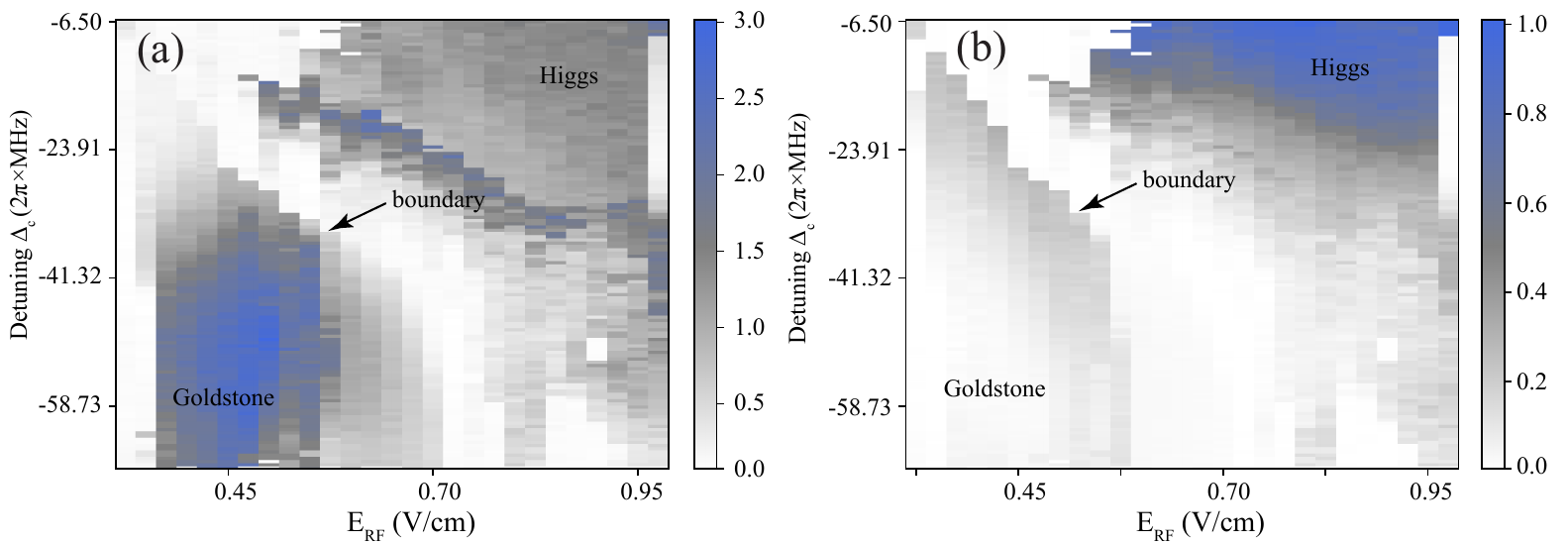}\\
\caption{\textbf{Variance spectrum.} (a) The variance of the phase $\phi$ versus the amplitude of the RF electric field $E_{\rm{RF}}$ and the coupling detuning $\Delta_c$. It displays distinct variance stages of Var($\phi$), with varying color contrasts in gray and blue, that indicate Higgs and Goldstone modes within the spectra. (b) The variance of the amplitude $\psi$ versus the amplitude of the RF electric field $E_{\rm{RF}}$ and the coupling detuning $\Delta_c$. The distinct variance stages of Var($\phi$) and Var($\psi$), with varying color contrasts display the boundaries [marked by the arrows] of U(1)-symmetry breaking.}
\label{Energy gap}
\end{figure*}

\subsection*{Variance spectrum}
The distinct phase variances Var($\phi$) and Var($|\psi|$) between Higgs and Goldstone modes enable us to distinguish them within variance spectrum. By varying the amplitude of the RF electric field $E_{\rm{RF}}$, we measure the variance of the phase $\phi$ and amplitude $|\psi|$, as illustrated in Fig.~\ref{Energy gap}(a) and Fig.~\ref{Energy gap}(b). In Figs.~\ref{Energy gap}(a) and (b), Higgs and Goldstone modes are evident, as indicated by distinct phase and amplitude variances. In the variance spectrum, the large Var($\phi$) (blue area in Fig.~\ref{Energy gap}(a)) and small Var($|\psi|$) observed around $E_{\rm{RF}}=0.45$ V/cm and $\Delta = -2\pi\times50$ MHz suggest the existence of a zero-energy state in the long-wavelength limit, i.e., along the azimuthal direction at the potential minimum in Fig.~\ref{setup}(c), corresponding to massless Goldstone mode. According to the Goldstone theorem \cite{goldstone1961field,goldstone1962broken}, the Goldstone mode exhibits a linear dispersion relation, where energy increases with momentum $q$. For long-wavelength (small $q$) Goldstone modes, the energy approaches zero \cite{pekker2015amplitude}. 

Meanwhile, the relatively small Var($\phi$) and large Var($|\psi|$) indicate the presence of the Higgs mode in the spectrum, which occurs at small detuning $|\Delta|$ and large $E_{\rm{RF}}$. The Higgs mode has a non-zero energy starting point because of the massive dispersion, leading to a distinct energy gap between the Higgs and Goldstone modes. According to the different masses $m_{\text{Higgs}}=\sqrt{4\hbar N(g^2N/(\Delta-\delta)-\delta)}$ and $m_{\text{Goldstone}}=0$, the Higgs mode shifts to smaller detuning $|\Delta|$ and is obvious to be gaped from Goldstone mode within relatively larger detuning $|\Delta|$. In Figs.~\ref{Energy gap}(a) and (b), we also observe a sharp boundary in the spectrum, indicated by a sudden change in color, which marks the onset of U(1) symmetry breaking. This occurs when the coefficient $\lambda$ changes from negative to positive as we vary $E_\text{RF}$ or $\Delta_c$. This effect leads to a gaped region along the diagonal, dividing the spectrum into an upper right and a lower left regimes, which is determined by the U(1) symmetry condition at $\lambda<0$ [affected by the unstable term $\hbar g^2 N^2/(\Delta-\delta)$ in coefficient $\lambda$ at the divergence $\Delta-\delta\rightarrow0$]. This behavior is consistent with the theoretical predictions outlined in the Methods section. The appearance of modes in the lower left region is attributed to the van der Waals interaction between Rydberg atoms. Due to the many-body interaction, more mass can be obtained by considering a lower left regime in Fig.~\ref{Energy gap}(b).

\section*{Discussions}
In our experiment, the Higgs and Goldstone modes are not completely separated in energy spectrum due to the Doppler distributed atoms, resulting in a degenerate region where these modes are indistinguishable in the energy spectrum. This is different from the coupling between amplitude and phase induced by particle-hole asymmetry \cite{pekker2015amplitude}. The study of the nontrivial aspects of Higgs and Goldstone modes in the context of thermal Rydberg gases has opened new avenues for exploring the non-equilibrium dynamics of collective modes and their signatures in room temperature.

These investigations are crucial for understanding symmetry breaking, offering valuable insights for both quantum simulations in high-energy physics and cosmology. The system of interacting Rydberg atoms can be considered analogous to the large number of interacting matter particles in galaxy clusters. In this analogy, the many-body interactions within the Rydberg atom ensemble generate additional Higgs and Goldstone modes, leading to the generation of additional mass. It opens pathways for exploring collective phenomena and additional mass generation, providing potential applications for connecting quantum systems to astrophysical observations of additional mass (dark matter) beyond ordinary matter.

In summary, we have observed both the Higgs and Goldstone modes in thermal interacting Rydberg gases as a result of U(1) symmetry breaking. The energy gap between these modes can be tuned by varying the external field, and thus the mass of the Higgs mode is modified. The emergence of amplitude and phase fluctuations plays a key role in understanding the stability and dynamics of the ordered phase, shedding light on how symmetry breaking influences collective behavior in Rydberg many-body systems. Ultimately, these findings deepen our understanding of how the Higgs mechanism functions across various physical contexts, especially including many-body interaction, bridging concepts from condensed matter physics to high-energy physics.

\bibliography{ref}

\begin{thebibliography}{43}%
\makeatletter
\providecommand \@ifxundefined [1]{%
 \@ifx{#1\undefined}
}%
\providecommand \@ifnum [1]{%
 \ifnum #1\expandafter \@firstoftwo
 \else \expandafter \@secondoftwo
 \fi
}%
\providecommand \@ifx [1]{%
 \ifx #1\expandafter \@firstoftwo
 \else \expandafter \@secondoftwo
 \fi
}%
\providecommand \natexlab [1]{#1}%
\providecommand \enquote  [1]{``#1''}%
\providecommand \bibnamefont  [1]{#1}%
\providecommand \bibfnamefont [1]{#1}%
\providecommand \citenamefont [1]{#1}%
\providecommand \href@noop [0]{\@secondoftwo}%
\providecommand \href [0]{\begingroup \@sanitize@url \@href}%
\providecommand \@href[1]{\@@startlink{#1}\@@href}%
\providecommand \@@href[1]{\endgroup#1\@@endlink}%
\providecommand \@sanitize@url [0]{\catcode `\\12\catcode `\$12\catcode
  `\&12\catcode `\#12\catcode `\^12\catcode `\_12\catcode `\%12\relax}%
\providecommand \@@startlink[1]{}%
\providecommand \@@endlink[0]{}%
\providecommand \url  [0]{\begingroup\@sanitize@url \@url }%
\providecommand \@url [1]{\endgroup\@href {#1}{\urlprefix }}%
\providecommand \urlprefix  [0]{URL }%
\providecommand \Eprint [0]{\href }%
\providecommand \doibase [0]{https://doi.org/}%
\providecommand \selectlanguage [0]{\@gobble}%
\providecommand \bibinfo  [0]{\@secondoftwo}%
\providecommand \bibfield  [0]{\@secondoftwo}%
\providecommand \translation [1]{[#1]}%
\providecommand \BibitemOpen [0]{}%
\providecommand \bibitemStop [0]{}%
\providecommand \bibitemNoStop [0]{.\EOS\space}%
\providecommand \EOS [0]{\spacefactor3000\relax}%
\providecommand \BibitemShut  [1]{\csname bibitem#1\endcsname}%
\let\auto@bib@innerbib\@empty
\bibitem [{\citenamefont {Goldstone}(1961)}]{goldstone1961field}%
  \BibitemOpen
  \bibfield  {author} {\bibinfo {author} {\bibfnamefont {J.}~\bibnamefont
  {Goldstone}},\ }\bibfield  {title} {\bibinfo {title} {Field theories with
  {\guillemotleft}superconductor{\guillemotright} solutions},\ }\href
  {https://link.springer.com/article/10.1007/BF02812722} {\bibfield  {journal}
  {\bibinfo  {journal} {Il Nuovo Cimento (1955-1965)}\ }\textbf {\bibinfo
  {volume} {19}},\ \bibinfo {pages} {154} (\bibinfo {year} {1961})}\BibitemShut
  {NoStop}%
\bibitem [{\citenamefont {Nambu}\ and\ \citenamefont
  {Jona-Lasinio}(1961)}]{nambu1961dynamical}%
  \BibitemOpen
  \bibfield  {author} {\bibinfo {author} {\bibfnamefont {Y.}~\bibnamefont
  {Nambu}}\ and\ \bibinfo {author} {\bibfnamefont {G.}~\bibnamefont
  {Jona-Lasinio}},\ }\bibfield  {title} {\bibinfo {title} {Dynamical model of
  elementary particles based on an analogy with superconductivity. i},\ }\href
  {https://journals.aps.org/pr/abstract/10.1103/PhysRev.122.345} {\bibfield
  {journal} {\bibinfo  {journal} {Physical review}\ }\textbf {\bibinfo {volume}
  {122}},\ \bibinfo {pages} {345} (\bibinfo {year} {1961})}\BibitemShut
  {NoStop}%
\bibitem [{\citenamefont {Goldstone}\ \emph {et~al.}(1962)\citenamefont
  {Goldstone}, \citenamefont {Salam},\ and\ \citenamefont
  {Weinberg}}]{goldstone1962broken}%
  \BibitemOpen
  \bibfield  {author} {\bibinfo {author} {\bibfnamefont {J.}~\bibnamefont
  {Goldstone}}, \bibinfo {author} {\bibfnamefont {A.}~\bibnamefont {Salam}},\
  and\ \bibinfo {author} {\bibfnamefont {S.}~\bibnamefont {Weinberg}},\
  }\bibfield  {title} {\bibinfo {title} {Broken symmetries},\ }\href
  {https://journals.aps.org/pr/abstract/10.1103/PhysRev.127.965} {\bibfield
  {journal} {\bibinfo  {journal} {Physical Review}\ }\textbf {\bibinfo {volume}
  {127}},\ \bibinfo {pages} {965} (\bibinfo {year} {1962})}\BibitemShut
  {NoStop}%
\bibitem [{\citenamefont {Anderson}(1963)}]{anderson1963plasmons}%
  \BibitemOpen
  \bibfield  {author} {\bibinfo {author} {\bibfnamefont {P.~W.}\ \bibnamefont
  {Anderson}},\ }\bibfield  {title} {\bibinfo {title} {Plasmons, gauge
  invariance, and mass},\ }\href
  {https://journals.aps.org/pr/abstract/10.1103/PhysRev.130.439} {\bibfield
  {journal} {\bibinfo  {journal} {Physical Review}\ }\textbf {\bibinfo {volume}
  {130}},\ \bibinfo {pages} {439} (\bibinfo {year} {1963})}\BibitemShut
  {NoStop}%
\bibitem [{\citenamefont {Englert}\ and\ \citenamefont
  {Brout}(1964)}]{englert1964broken}%
  \BibitemOpen
  \bibfield  {author} {\bibinfo {author} {\bibfnamefont {F.}~\bibnamefont
  {Englert}}\ and\ \bibinfo {author} {\bibfnamefont {R.}~\bibnamefont
  {Brout}},\ }\bibfield  {title} {\bibinfo {title} {Broken symmetry and the
  mass of gauge vector mesons},\ }\href
  {https://journals.aps.org/prl/abstract/10.1103/PhysRevLett.13.321} {\bibfield
   {journal} {\bibinfo  {journal} {Physical review letters}\ }\textbf {\bibinfo
  {volume} {13}},\ \bibinfo {pages} {321} (\bibinfo {year} {1964})}\BibitemShut
  {NoStop}%
\bibitem [{\citenamefont {Weinberg}(1967)}]{weinberg1967model}%
  \BibitemOpen
  \bibfield  {author} {\bibinfo {author} {\bibfnamefont {S.}~\bibnamefont
  {Weinberg}},\ }\bibfield  {title} {\bibinfo {title} {A model of leptons},\
  }\href {https://journals.aps.org/prl/abstract/10.1103/PhysRevLett.19.1264}
  {\bibfield  {journal} {\bibinfo  {journal} {Physical review letters}\
  }\textbf {\bibinfo {volume} {19}},\ \bibinfo {pages} {1264} (\bibinfo {year}
  {1967})}\BibitemShut {NoStop}%
\bibitem [{\citenamefont {Sooryakumar}\ and\ \citenamefont
  {Klein}(1980)}]{sooryakumar1980raman}%
  \BibitemOpen
  \bibfield  {author} {\bibinfo {author} {\bibfnamefont {R.}~\bibnamefont
  {Sooryakumar}}\ and\ \bibinfo {author} {\bibfnamefont {M.}~\bibnamefont
  {Klein}},\ }\bibfield  {title} {\bibinfo {title} {Raman scattering by
  superconducting-gap excitations and their coupling to charge-density waves},\
  }\href {https://journals.aps.org/prl/abstract/10.1103/PhysRevLett.45.660}
  {\bibfield  {journal} {\bibinfo  {journal} {Physical Review Letters}\
  }\textbf {\bibinfo {volume} {45}},\ \bibinfo {pages} {660} (\bibinfo {year}
  {1980})}\BibitemShut {NoStop}%
\bibitem [{\citenamefont {Littlewood}\ and\ \citenamefont
  {Varma}(1982)}]{littlewood1982amplitude}%
  \BibitemOpen
  \bibfield  {author} {\bibinfo {author} {\bibfnamefont {P.}~\bibnamefont
  {Littlewood}}\ and\ \bibinfo {author} {\bibfnamefont {C.}~\bibnamefont
  {Varma}},\ }\bibfield  {title} {\bibinfo {title} {Amplitude collective modes
  in superconductors and their coupling to charge-density waves},\ }\href
  {https://journals.aps.org/prb/abstract/10.1103/PhysRevB.26.4883} {\bibfield
  {journal} {\bibinfo  {journal} {Physical Review B}\ }\textbf {\bibinfo
  {volume} {26}},\ \bibinfo {pages} {4883} (\bibinfo {year}
  {1982})}\BibitemShut {NoStop}%
\bibitem [{\citenamefont {Varma}(2002)}]{varma2002higgs}%
  \BibitemOpen
  \bibfield  {author} {\bibinfo {author} {\bibfnamefont {C.}~\bibnamefont
  {Varma}},\ }\bibfield  {title} {\bibinfo {title} {Higgs boson in
  superconductors},\ }\href
  {https://link.springer.com/article/10.1023/A:1013890507658} {\bibfield
  {journal} {\bibinfo  {journal} {Journal of low temperature physics}\ }\textbf
  {\bibinfo {volume} {126}},\ \bibinfo {pages} {901} (\bibinfo {year}
  {2002})}\BibitemShut {NoStop}%
\bibitem [{\citenamefont {Pekker}\ and\ \citenamefont
  {Varma}(2015)}]{pekker2015amplitude}%
  \BibitemOpen
  \bibfield  {author} {\bibinfo {author} {\bibfnamefont {D.}~\bibnamefont
  {Pekker}}\ and\ \bibinfo {author} {\bibfnamefont {C.}~\bibnamefont {Varma}},\
  }\bibfield  {title} {\bibinfo {title} {Amplitude/higgs modes in condensed
  matter physics},\ }\href
  {https://www.annualreviews.org/content/journals/10.1146/annurev-conmatphys-031214-014350}
  {\bibfield  {journal} {\bibinfo  {journal} {Annu. Rev. Condens. Matter
  Phys.}\ }\textbf {\bibinfo {volume} {6}},\ \bibinfo {pages} {269} (\bibinfo
  {year} {2015})}\BibitemShut {NoStop}%
\bibitem [{\citenamefont {Strand}\ \emph {et~al.}(2010)\citenamefont {Strand},
  \citenamefont {Bahr}, \citenamefont {Van~Harlingen}, \citenamefont {Davis},
  \citenamefont {Gannon},\ and\ \citenamefont
  {Halperin}}]{strand2010transition}%
  \BibitemOpen
  \bibfield  {author} {\bibinfo {author} {\bibfnamefont {J.}~\bibnamefont
  {Strand}}, \bibinfo {author} {\bibfnamefont {D.}~\bibnamefont {Bahr}},
  \bibinfo {author} {\bibfnamefont {D.}~\bibnamefont {Van~Harlingen}}, \bibinfo
  {author} {\bibfnamefont {J.}~\bibnamefont {Davis}}, \bibinfo {author}
  {\bibfnamefont {W.}~\bibnamefont {Gannon}},\ and\ \bibinfo {author}
  {\bibfnamefont {W.}~\bibnamefont {Halperin}},\ }\bibfield  {title} {\bibinfo
  {title} {The transition between real and complex superconducting order
  parameter phases in upt3},\ }\href
  {https://www.science.org/doi/abs/10.1126/science.1187943} {\bibfield
  {journal} {\bibinfo  {journal} {Science}\ }\textbf {\bibinfo {volume}
  {328}},\ \bibinfo {pages} {1368} (\bibinfo {year} {2010})}\BibitemShut
  {NoStop}%
\bibitem [{\citenamefont {Matsunaga}\ \emph {et~al.}(2014)\citenamefont
  {Matsunaga}, \citenamefont {Tsuji}, \citenamefont {Fujita}, \citenamefont
  {Sugioka}, \citenamefont {Makise}, \citenamefont {Uzawa}, \citenamefont
  {Terai}, \citenamefont {Wang}, \citenamefont {Aoki},\ and\ \citenamefont
  {Shimano}}]{matsunaga2014light}%
  \BibitemOpen
  \bibfield  {author} {\bibinfo {author} {\bibfnamefont {R.}~\bibnamefont
  {Matsunaga}}, \bibinfo {author} {\bibfnamefont {N.}~\bibnamefont {Tsuji}},
  \bibinfo {author} {\bibfnamefont {H.}~\bibnamefont {Fujita}}, \bibinfo
  {author} {\bibfnamefont {A.}~\bibnamefont {Sugioka}}, \bibinfo {author}
  {\bibfnamefont {K.}~\bibnamefont {Makise}}, \bibinfo {author} {\bibfnamefont
  {Y.}~\bibnamefont {Uzawa}}, \bibinfo {author} {\bibfnamefont
  {H.}~\bibnamefont {Terai}}, \bibinfo {author} {\bibfnamefont
  {Z.}~\bibnamefont {Wang}}, \bibinfo {author} {\bibfnamefont {H.}~\bibnamefont
  {Aoki}},\ and\ \bibinfo {author} {\bibfnamefont {R.}~\bibnamefont
  {Shimano}},\ }\bibfield  {title} {\bibinfo {title} {Light-induced collective
  pseudospin precession resonating with higgs mode in a superconductor},\
  }\href {https://www.science.org/doi/abs/10.1126/science.1254697} {\bibfield
  {journal} {\bibinfo  {journal} {Science}\ }\textbf {\bibinfo {volume}
  {345}},\ \bibinfo {pages} {1145} (\bibinfo {year} {2014})}\BibitemShut
  {NoStop}%
\bibitem [{\citenamefont {Sherman}\ \emph {et~al.}(2015)\citenamefont
  {Sherman}, \citenamefont {Pracht}, \citenamefont {Gorshunov}, \citenamefont
  {Poran}, \citenamefont {Jesudasan}, \citenamefont {Chand}, \citenamefont
  {Raychaudhuri}, \citenamefont {Swanson}, \citenamefont {Trivedi},
  \citenamefont {Auerbach} \emph {et~al.}}]{sherman2015higgs}%
  \BibitemOpen
  \bibfield  {author} {\bibinfo {author} {\bibfnamefont {D.}~\bibnamefont
  {Sherman}}, \bibinfo {author} {\bibfnamefont {U.~S.}\ \bibnamefont {Pracht}},
  \bibinfo {author} {\bibfnamefont {B.}~\bibnamefont {Gorshunov}}, \bibinfo
  {author} {\bibfnamefont {S.}~\bibnamefont {Poran}}, \bibinfo {author}
  {\bibfnamefont {J.}~\bibnamefont {Jesudasan}}, \bibinfo {author}
  {\bibfnamefont {M.}~\bibnamefont {Chand}}, \bibinfo {author} {\bibfnamefont
  {P.}~\bibnamefont {Raychaudhuri}}, \bibinfo {author} {\bibfnamefont
  {M.}~\bibnamefont {Swanson}}, \bibinfo {author} {\bibfnamefont
  {N.}~\bibnamefont {Trivedi}}, \bibinfo {author} {\bibfnamefont
  {A.}~\bibnamefont {Auerbach}}, \emph {et~al.},\ }\bibfield  {title} {\bibinfo
  {title} {The higgs mode in disordered superconductors close to a quantum
  phase transition},\ }\href {https://www.nature.com/articles/nphys3227}
  {\bibfield  {journal} {\bibinfo  {journal} {Nature Physics}\ }\textbf
  {\bibinfo {volume} {11}},\ \bibinfo {pages} {188} (\bibinfo {year}
  {2015})}\BibitemShut {NoStop}%
\bibitem [{\citenamefont {Pollet}\ and\ \citenamefont
  {Prokof’Ev}(2012)}]{pollet2012higgs}%
  \BibitemOpen
  \bibfield  {author} {\bibinfo {author} {\bibfnamefont {L.}~\bibnamefont
  {Pollet}}\ and\ \bibinfo {author} {\bibfnamefont {N.}~\bibnamefont
  {Prokof’Ev}},\ }\bibfield  {title} {\bibinfo {title} {Higgs mode in a
  two-dimensional superfluid},\ }\href
  {https://journals.aps.org/prl/abstract/10.1103/PhysRevLett.109.010401}
  {\bibfield  {journal} {\bibinfo  {journal} {Physical Review Letters}\
  }\textbf {\bibinfo {volume} {109}},\ \bibinfo {pages} {010401} (\bibinfo
  {year} {2012})}\BibitemShut {NoStop}%
\bibitem [{\citenamefont {Endres}\ \emph {et~al.}(2012)\citenamefont {Endres},
  \citenamefont {Fukuhara}, \citenamefont {Pekker}, \citenamefont {Cheneau},
  \citenamefont {Schau$\beta$}, \citenamefont {Gross}, \citenamefont {Demler},
  \citenamefont {Kuhr},\ and\ \citenamefont {Bloch}}]{endres2012higgs}%
  \BibitemOpen
  \bibfield  {author} {\bibinfo {author} {\bibfnamefont {M.}~\bibnamefont
  {Endres}}, \bibinfo {author} {\bibfnamefont {T.}~\bibnamefont {Fukuhara}},
  \bibinfo {author} {\bibfnamefont {D.}~\bibnamefont {Pekker}}, \bibinfo
  {author} {\bibfnamefont {M.}~\bibnamefont {Cheneau}}, \bibinfo {author}
  {\bibfnamefont {P.}~\bibnamefont {Schau$\beta$}}, \bibinfo {author}
  {\bibfnamefont {C.}~\bibnamefont {Gross}}, \bibinfo {author} {\bibfnamefont
  {E.}~\bibnamefont {Demler}}, \bibinfo {author} {\bibfnamefont
  {S.}~\bibnamefont {Kuhr}},\ and\ \bibinfo {author} {\bibfnamefont
  {I.}~\bibnamefont {Bloch}},\ }\bibfield  {title} {\bibinfo {title} {The
  ‘higgs’ amplitude mode at the two-dimensional superfluid/mott insulator
  transition},\ }\href {https://www.nature.com/articles/nature11255} {\bibfield
   {journal} {\bibinfo  {journal} {Nature}\ }\textbf {\bibinfo {volume}
  {487}},\ \bibinfo {pages} {454} (\bibinfo {year} {2012})}\BibitemShut
  {NoStop}%
\bibitem [{\citenamefont {L{\'e}onard}\ \emph {et~al.}(2017)\citenamefont
  {L{\'e}onard}, \citenamefont {Morales}, \citenamefont {Zupancic},
  \citenamefont {Donner},\ and\ \citenamefont
  {Esslinger}}]{leonard2017monitoring}%
  \BibitemOpen
  \bibfield  {author} {\bibinfo {author} {\bibfnamefont {J.}~\bibnamefont
  {L{\'e}onard}}, \bibinfo {author} {\bibfnamefont {A.}~\bibnamefont
  {Morales}}, \bibinfo {author} {\bibfnamefont {P.}~\bibnamefont {Zupancic}},
  \bibinfo {author} {\bibfnamefont {T.}~\bibnamefont {Donner}},\ and\ \bibinfo
  {author} {\bibfnamefont {T.}~\bibnamefont {Esslinger}},\ }\bibfield  {title}
  {\bibinfo {title} {Monitoring and manipulating higgs and goldstone modes in a
  supersolid quantum gas},\ }\href
  {https://www.science.org/doi/full/10.1126/science.aan2608} {\bibfield
  {journal} {\bibinfo  {journal} {Science}\ }\textbf {\bibinfo {volume}
  {358}},\ \bibinfo {pages} {1415} (\bibinfo {year} {2017})}\BibitemShut
  {NoStop}%
\bibitem [{\citenamefont {Behrle}\ \emph {et~al.}(2018)\citenamefont {Behrle},
  \citenamefont {Harrison}, \citenamefont {Kombe}, \citenamefont {Gao},
  \citenamefont {Link}, \citenamefont {Bernier}, \citenamefont {Kollath},\ and\
  \citenamefont {K{\"o}hl}}]{behrle2018higgs}%
  \BibitemOpen
  \bibfield  {author} {\bibinfo {author} {\bibfnamefont {A.}~\bibnamefont
  {Behrle}}, \bibinfo {author} {\bibfnamefont {T.}~\bibnamefont {Harrison}},
  \bibinfo {author} {\bibfnamefont {J.}~\bibnamefont {Kombe}}, \bibinfo
  {author} {\bibfnamefont {K.}~\bibnamefont {Gao}}, \bibinfo {author}
  {\bibfnamefont {M.}~\bibnamefont {Link}}, \bibinfo {author} {\bibfnamefont
  {J.-S.}\ \bibnamefont {Bernier}}, \bibinfo {author} {\bibfnamefont
  {C.}~\bibnamefont {Kollath}},\ and\ \bibinfo {author} {\bibfnamefont
  {M.}~\bibnamefont {K{\"o}hl}},\ }\bibfield  {title} {\bibinfo {title} {Higgs
  mode in a strongly interacting fermionic superfluid},\ }\href
  {https://www.nature.com/articles/s41567-018-0128-6} {\bibfield  {journal}
  {\bibinfo  {journal} {Nature Physics}\ }\textbf {\bibinfo {volume} {14}},\
  \bibinfo {pages} {781} (\bibinfo {year} {2018})}\BibitemShut {NoStop}%
\bibitem [{\citenamefont {Guo}\ \emph {et~al.}(2019)\citenamefont {Guo},
  \citenamefont {B{\"o}ttcher}, \citenamefont {Hertkorn}, \citenamefont
  {Schmidt}, \citenamefont {Wenzel}, \citenamefont {B{\"u}chler}, \citenamefont
  {Langen},\ and\ \citenamefont {Pfau}}]{guo2019low}%
  \BibitemOpen
  \bibfield  {author} {\bibinfo {author} {\bibfnamefont {M.}~\bibnamefont
  {Guo}}, \bibinfo {author} {\bibfnamefont {F.}~\bibnamefont {B{\"o}ttcher}},
  \bibinfo {author} {\bibfnamefont {J.}~\bibnamefont {Hertkorn}}, \bibinfo
  {author} {\bibfnamefont {J.-N.}\ \bibnamefont {Schmidt}}, \bibinfo {author}
  {\bibfnamefont {M.}~\bibnamefont {Wenzel}}, \bibinfo {author} {\bibfnamefont
  {H.~P.}\ \bibnamefont {B{\"u}chler}}, \bibinfo {author} {\bibfnamefont
  {T.}~\bibnamefont {Langen}},\ and\ \bibinfo {author} {\bibfnamefont
  {T.}~\bibnamefont {Pfau}},\ }\bibfield  {title} {\bibinfo {title} {The
  low-energy goldstone mode in a trapped dipolar supersolid},\ }\href
  {https://www.nature.com/articles/s41586-019-1569-5} {\bibfield  {journal}
  {\bibinfo  {journal} {Nature}\ }\textbf {\bibinfo {volume} {574}},\ \bibinfo
  {pages} {386} (\bibinfo {year} {2019})}\BibitemShut {NoStop}%
\bibitem [{\citenamefont {Geier}\ \emph {et~al.}(2021)\citenamefont {Geier},
  \citenamefont {Martone}, \citenamefont {Hauke},\ and\ \citenamefont
  {Stringari}}]{geier2021exciting}%
  \BibitemOpen
  \bibfield  {author} {\bibinfo {author} {\bibfnamefont {K.~T.}\ \bibnamefont
  {Geier}}, \bibinfo {author} {\bibfnamefont {G.~I.}\ \bibnamefont {Martone}},
  \bibinfo {author} {\bibfnamefont {P.}~\bibnamefont {Hauke}},\ and\ \bibinfo
  {author} {\bibfnamefont {S.}~\bibnamefont {Stringari}},\ }\bibfield  {title}
  {\bibinfo {title} {Exciting the goldstone modes of a supersolid
  spin-orbit-coupled bose gas},\ }\href
  {https://journals.aps.org/prl/abstract/10.1103/PhysRevLett.127.115301}
  {\bibfield  {journal} {\bibinfo  {journal} {Physical Review Letters}\
  }\textbf {\bibinfo {volume} {127}},\ \bibinfo {pages} {115301} (\bibinfo
  {year} {2021})}\BibitemShut {NoStop}%
\bibitem [{\citenamefont {Hong}\ \emph {et~al.}(2017)\citenamefont {Hong},
  \citenamefont {Matsumoto}, \citenamefont {Qiu}, \citenamefont {Chen},
  \citenamefont {Gentile}, \citenamefont {Watson}, \citenamefont {Awwadi},
  \citenamefont {Turnbull}, \citenamefont {Dissanayake}, \citenamefont
  {Agrawal} \emph {et~al.}}]{hong2017higgs}%
  \BibitemOpen
  \bibfield  {author} {\bibinfo {author} {\bibfnamefont {T.}~\bibnamefont
  {Hong}}, \bibinfo {author} {\bibfnamefont {M.}~\bibnamefont {Matsumoto}},
  \bibinfo {author} {\bibfnamefont {Y.}~\bibnamefont {Qiu}}, \bibinfo {author}
  {\bibfnamefont {W.}~\bibnamefont {Chen}}, \bibinfo {author} {\bibfnamefont
  {T.~R.}\ \bibnamefont {Gentile}}, \bibinfo {author} {\bibfnamefont
  {S.}~\bibnamefont {Watson}}, \bibinfo {author} {\bibfnamefont {F.~F.}\
  \bibnamefont {Awwadi}}, \bibinfo {author} {\bibfnamefont {M.~M.}\
  \bibnamefont {Turnbull}}, \bibinfo {author} {\bibfnamefont {S.~E.}\
  \bibnamefont {Dissanayake}}, \bibinfo {author} {\bibfnamefont
  {H.}~\bibnamefont {Agrawal}}, \emph {et~al.},\ }\bibfield  {title} {\bibinfo
  {title} {Higgs amplitude mode in a two-dimensional quantum antiferromagnet
  near the quantum critical point},\ }\href
  {https://www.nature.com/articles/nphys4182} {\bibfield  {journal} {\bibinfo
  {journal} {Nature Physics}\ }\textbf {\bibinfo {volume} {13}},\ \bibinfo
  {pages} {638} (\bibinfo {year} {2017})}\BibitemShut {NoStop}%
\bibitem [{\citenamefont {Saffman}\ \emph {et~al.}(2010)\citenamefont
  {Saffman}, \citenamefont {Walker},\ and\ \citenamefont
  {M\o{}lmer}}]{saffman2010quantum}%
  \BibitemOpen
  \bibfield  {author} {\bibinfo {author} {\bibfnamefont {M.}~\bibnamefont
  {Saffman}}, \bibinfo {author} {\bibfnamefont {T.~G.}\ \bibnamefont
  {Walker}},\ and\ \bibinfo {author} {\bibfnamefont {K.}~\bibnamefont
  {M\o{}lmer}},\ }\bibfield  {title} {\bibinfo {title} {Quantum information
  with {Rydberg} atoms},\ }\href {https://doi.org/10.1103/RevModPhys.82.2313}
  {\bibfield  {journal} {\bibinfo  {journal} {Rev. Mod. Phys.}\ }\textbf
  {\bibinfo {volume} {82}},\ \bibinfo {pages} {2313} (\bibinfo {year}
  {2010})}\BibitemShut {NoStop}%
\bibitem [{\citenamefont {Adams}\ \emph {et~al.}(2019)\citenamefont {Adams},
  \citenamefont {Pritchard},\ and\ \citenamefont {Shaffer}}]{adams2019rydberg}%
  \BibitemOpen
  \bibfield  {author} {\bibinfo {author} {\bibfnamefont {C.~S.}\ \bibnamefont
  {Adams}}, \bibinfo {author} {\bibfnamefont {J.~D.}\ \bibnamefont
  {Pritchard}},\ and\ \bibinfo {author} {\bibfnamefont {J.~P.}\ \bibnamefont
  {Shaffer}},\ }\bibfield  {title} {\bibinfo {title} {Rydberg atom quantum
  technologies},\ }\href {https://doi.org/10.1088/1361-6455/ab52ef} {\bibfield
  {journal} {\bibinfo  {journal} {J. Phys. B: At. Mol. Opt. Phys.}\ }\textbf
  {\bibinfo {volume} {53}},\ \bibinfo {pages} {012002} (\bibinfo {year}
  {2019})}\BibitemShut {NoStop}%
\bibitem [{\citenamefont {Browaeys}\ and\ \citenamefont
  {Lahaye}(2020)}]{browaeys2020many}%
  \BibitemOpen
  \bibfield  {author} {\bibinfo {author} {\bibfnamefont {A.}~\bibnamefont
  {Browaeys}}\ and\ \bibinfo {author} {\bibfnamefont {T.}~\bibnamefont
  {Lahaye}},\ }\bibfield  {title} {\bibinfo {title} {Many-body physics with
  individually controlled {R}ydberg atoms},\ }\href
  {https://doi.org/10.1038/s41567-019-0733-z} {\bibfield  {journal} {\bibinfo
  {journal} {Nat. Phys.}\ }\textbf {\bibinfo {volume} {16}},\ \bibinfo {pages}
  {132} (\bibinfo {year} {2020})}\BibitemShut {NoStop}%
\bibitem [{\citenamefont {Lee}\ \emph {et~al.}(2012)\citenamefont {Lee},
  \citenamefont {H\"affner},\ and\ \citenamefont {Cross}}]{lee2012collective}%
  \BibitemOpen
  \bibfield  {author} {\bibinfo {author} {\bibfnamefont {T.~E.}\ \bibnamefont
  {Lee}}, \bibinfo {author} {\bibfnamefont {H.}~\bibnamefont {H\"affner}},\
  and\ \bibinfo {author} {\bibfnamefont {M.~C.}\ \bibnamefont {Cross}},\
  }\bibfield  {title} {\bibinfo {title} {Collective quantum jumps of {Rydberg}
  atoms},\ }\href {https://doi.org/10.1103/PhysRevLett.108.023602} {\bibfield
  {journal} {\bibinfo  {journal} {Phys. Rev. Lett.}\ }\textbf {\bibinfo
  {volume} {108}},\ \bibinfo {pages} {023602} (\bibinfo {year}
  {2012})}\BibitemShut {NoStop}%
\bibitem [{\citenamefont {Carr}\ \emph {et~al.}(2013)\citenamefont {Carr},
  \citenamefont {Ritter}, \citenamefont {Wade}, \citenamefont {Adams},\ and\
  \citenamefont {Weatherill}}]{carr2013nonequilibrium}%
  \BibitemOpen
  \bibfield  {author} {\bibinfo {author} {\bibfnamefont {C.}~\bibnamefont
  {Carr}}, \bibinfo {author} {\bibfnamefont {R.}~\bibnamefont {Ritter}},
  \bibinfo {author} {\bibfnamefont {C.}~\bibnamefont {Wade}}, \bibinfo {author}
  {\bibfnamefont {C.~S.}\ \bibnamefont {Adams}},\ and\ \bibinfo {author}
  {\bibfnamefont {K.~J.}\ \bibnamefont {Weatherill}},\ }\bibfield  {title}
  {\bibinfo {title} {Nonequilibrium phase transition in a dilute {Rydberg}
  ensemble},\ }\href
  {https://doi.org/https://doi.org/10.1103/PhysRevLett.111.113901} {\bibfield
  {journal} {\bibinfo  {journal} {Phys. Rev. Lett.}\ }\textbf {\bibinfo
  {volume} {111}},\ \bibinfo {pages} {113901} (\bibinfo {year}
  {2013})}\BibitemShut {NoStop}%
\bibitem [{\citenamefont {Schempp}\ \emph {et~al.}(2014)\citenamefont
  {Schempp}, \citenamefont {G\"unter}, \citenamefont {Robert-de Saint-Vincent},
  \citenamefont {Hofmann}, \citenamefont {Breyel}, \citenamefont {Komnik},
  \citenamefont {Sch\"onleber}, \citenamefont {G\"arttner}, \citenamefont
  {Evers}, \citenamefont {Whitlock},\ and\ \citenamefont
  {Weidem\"uller}}]{schempp2014full}%
  \BibitemOpen
  \bibfield  {author} {\bibinfo {author} {\bibfnamefont {H.}~\bibnamefont
  {Schempp}}, \bibinfo {author} {\bibfnamefont {G.}~\bibnamefont {G\"unter}},
  \bibinfo {author} {\bibfnamefont {M.}~\bibnamefont {Robert-de
  Saint-Vincent}}, \bibinfo {author} {\bibfnamefont {C.~S.}\ \bibnamefont
  {Hofmann}}, \bibinfo {author} {\bibfnamefont {D.}~\bibnamefont {Breyel}},
  \bibinfo {author} {\bibfnamefont {A.}~\bibnamefont {Komnik}}, \bibinfo
  {author} {\bibfnamefont {D.~W.}\ \bibnamefont {Sch\"onleber}}, \bibinfo
  {author} {\bibfnamefont {M.}~\bibnamefont {G\"arttner}}, \bibinfo {author}
  {\bibfnamefont {J.}~\bibnamefont {Evers}}, \bibinfo {author} {\bibfnamefont
  {S.}~\bibnamefont {Whitlock}},\ and\ \bibinfo {author} {\bibfnamefont
  {M.}~\bibnamefont {Weidem\"uller}},\ }\bibfield  {title} {\bibinfo {title}
  {Full counting statistics of laser excited {Rydberg} aggregates in a
  one-dimensional geometry},\ }\href
  {https://doi.org/10.1103/PhysRevLett.112.013002} {\bibfield  {journal}
  {\bibinfo  {journal} {Phys. Rev. Lett.}\ }\textbf {\bibinfo {volume} {112}},\
  \bibinfo {pages} {013002} (\bibinfo {year} {2014})}\BibitemShut {NoStop}%
\bibitem [{\citenamefont {Marcuzzi}\ \emph {et~al.}(2014)\citenamefont
  {Marcuzzi}, \citenamefont {Levi}, \citenamefont {Diehl}, \citenamefont
  {Garrahan},\ and\ \citenamefont {Lesanovsky}}]{marcuzzi2014universal}%
  \BibitemOpen
  \bibfield  {author} {\bibinfo {author} {\bibfnamefont {M.}~\bibnamefont
  {Marcuzzi}}, \bibinfo {author} {\bibfnamefont {E.}~\bibnamefont {Levi}},
  \bibinfo {author} {\bibfnamefont {S.}~\bibnamefont {Diehl}}, \bibinfo
  {author} {\bibfnamefont {J.~P.}\ \bibnamefont {Garrahan}},\ and\ \bibinfo
  {author} {\bibfnamefont {I.}~\bibnamefont {Lesanovsky}},\ }\bibfield  {title}
  {\bibinfo {title} {Universal nonequilibrium properties of dissipative
  {Rydberg} gases},\ }\href
  {https://doi.org/https://doi.org/10.1103/PhysRevLett.113.210401} {\bibfield
  {journal} {\bibinfo  {journal} {Phys. Rev. Lett.}\ }\textbf {\bibinfo
  {volume} {113}},\ \bibinfo {pages} {210401} (\bibinfo {year}
  {2014})}\BibitemShut {NoStop}%
\bibitem [{\citenamefont {Lesanovsky}\ and\ \citenamefont
  {Garrahan}(2014)}]{lesanovsky2014out}%
  \BibitemOpen
  \bibfield  {author} {\bibinfo {author} {\bibfnamefont {I.}~\bibnamefont
  {Lesanovsky}}\ and\ \bibinfo {author} {\bibfnamefont {J.~P.}\ \bibnamefont
  {Garrahan}},\ }\bibfield  {title} {\bibinfo {title} {Out-of-equilibrium
  structures in strongly interacting {R}ydberg gases with dissipation},\ }\href
  {https://doi.org/10.1103/PhysRevA.90.011603} {\bibfield  {journal} {\bibinfo
  {journal} {Phys. Rev. A}\ }\textbf {\bibinfo {volume} {90}},\ \bibinfo
  {pages} {011603} (\bibinfo {year} {2014})}\BibitemShut {NoStop}%
\bibitem [{\citenamefont {Urvoy}\ \emph {et~al.}(2015)\citenamefont {Urvoy},
  \citenamefont {Ripka}, \citenamefont {Lesanovsky}, \citenamefont {Booth},
  \citenamefont {Shaffer}, \citenamefont {Pfau},\ and\ \citenamefont
  {L{\"o}w}}]{urvoy2015strongly}%
  \BibitemOpen
  \bibfield  {author} {\bibinfo {author} {\bibfnamefont {A.}~\bibnamefont
  {Urvoy}}, \bibinfo {author} {\bibfnamefont {F.}~\bibnamefont {Ripka}},
  \bibinfo {author} {\bibfnamefont {I.}~\bibnamefont {Lesanovsky}}, \bibinfo
  {author} {\bibfnamefont {D.}~\bibnamefont {Booth}}, \bibinfo {author}
  {\bibfnamefont {J.}~\bibnamefont {Shaffer}}, \bibinfo {author} {\bibfnamefont
  {T.}~\bibnamefont {Pfau}},\ and\ \bibinfo {author} {\bibfnamefont
  {R.}~\bibnamefont {L{\"o}w}},\ }\bibfield  {title} {\bibinfo {title}
  {Strongly correlated growth of {Rydberg} aggregates in a vapor cell},\ }\href
  {https://journals.aps.org/prl/abstract/10.1103/PhysRevLett.114.203002}
  {\bibfield  {journal} {\bibinfo  {journal} {Phys. Rev. Lett.}\ }\textbf
  {\bibinfo {volume} {114}},\ \bibinfo {pages} {203002} (\bibinfo {year}
  {2015})}\BibitemShut {NoStop}%
\bibitem [{\citenamefont {Ding}\ \emph {et~al.}(2022)\citenamefont {Ding},
  \citenamefont {Liu}, \citenamefont {Shi}, \citenamefont {Guo}, \citenamefont
  {M{\o}lmer},\ and\ \citenamefont {Adams}}]{ding2022enhanced}%
  \BibitemOpen
  \bibfield  {author} {\bibinfo {author} {\bibfnamefont {D.-S.}\ \bibnamefont
  {Ding}}, \bibinfo {author} {\bibfnamefont {Z.-K.}\ \bibnamefont {Liu}},
  \bibinfo {author} {\bibfnamefont {B.-S.}\ \bibnamefont {Shi}}, \bibinfo
  {author} {\bibfnamefont {G.-C.}\ \bibnamefont {Guo}}, \bibinfo {author}
  {\bibfnamefont {K.}~\bibnamefont {M{\o}lmer}},\ and\ \bibinfo {author}
  {\bibfnamefont {C.~S.}\ \bibnamefont {Adams}},\ }\bibfield  {title} {\bibinfo
  {title} {Enhanced metrology at the critical point of a many-body {Rydberg}
  atomic system},\ }\href {https://www.nature.com/articles/s41567-022-01777-8}
  {\bibfield  {journal} {\bibinfo  {journal} {Nat. Phys.}\ }\textbf {\bibinfo
  {volume} {18}},\ \bibinfo {pages} {1447} (\bibinfo {year}
  {2022})}\BibitemShut {NoStop}%
\bibitem [{\citenamefont {Helmrich}\ \emph {et~al.}(2020)\citenamefont
  {Helmrich}, \citenamefont {Arias}, \citenamefont {Lochead}, \citenamefont
  {Wintermantel}, \citenamefont {Buchhold}, \citenamefont {Diehl},\ and\
  \citenamefont {Whitlock}}]{Signatures2020Helmrich}%
  \BibitemOpen
  \bibfield  {author} {\bibinfo {author} {\bibfnamefont {S.}~\bibnamefont
  {Helmrich}}, \bibinfo {author} {\bibfnamefont {A.}~\bibnamefont {Arias}},
  \bibinfo {author} {\bibfnamefont {G.}~\bibnamefont {Lochead}}, \bibinfo
  {author} {\bibfnamefont {T.~M.}\ \bibnamefont {Wintermantel}}, \bibinfo
  {author} {\bibfnamefont {M.}~\bibnamefont {Buchhold}}, \bibinfo {author}
  {\bibfnamefont {S.}~\bibnamefont {Diehl}},\ and\ \bibinfo {author}
  {\bibfnamefont {S.}~\bibnamefont {Whitlock}},\ }\bibfield  {title} {\bibinfo
  {title} {Signatures of self-organized criticality in an ultracold atomic
  gas},\ }\href {https://doi.org/10.1038/s41586-019-1908-6} {\bibfield
  {journal} {\bibinfo  {journal} {Nature}\ }\textbf {\bibinfo {volume} {577}},\
  \bibinfo {pages} {481} (\bibinfo {year} {2020})}\BibitemShut {NoStop}%
\bibitem [{\citenamefont {Ding}\ \emph {et~al.}(2020)\citenamefont {Ding},
  \citenamefont {Busche}, \citenamefont {Shi}, \citenamefont {Guo},\ and\
  \citenamefont {Adams}}]{ding2019Phase}%
  \BibitemOpen
  \bibfield  {author} {\bibinfo {author} {\bibfnamefont {D.-S.}\ \bibnamefont
  {Ding}}, \bibinfo {author} {\bibfnamefont {H.}~\bibnamefont {Busche}},
  \bibinfo {author} {\bibfnamefont {B.-S.}\ \bibnamefont {Shi}}, \bibinfo
  {author} {\bibfnamefont {G.-C.}\ \bibnamefont {Guo}},\ and\ \bibinfo {author}
  {\bibfnamefont {C.~S.}\ \bibnamefont {Adams}},\ }\bibfield  {title} {\bibinfo
  {title} {Phase diagram of non-equilibrium phase transition in a
  strongly-interacting {Rydberg} atom vapour},\ }\href
  {https://journals.aps.org/prx/abstract/10.1103/PhysRevX.10.021023} {\bibfield
   {journal} {\bibinfo  {journal} {Phys. Rev. X}\ }\textbf {\bibinfo {volume}
  {10}},\ \bibinfo {pages} {021023} (\bibinfo {year} {2020})}\BibitemShut
  {NoStop}%
\bibitem [{\citenamefont {Wintermantel}\ \emph {et~al.}(2020)\citenamefont
  {Wintermantel}, \citenamefont {Wang}, \citenamefont {Lochead}, \citenamefont
  {Shevate}, \citenamefont {Brennen},\ and\ \citenamefont
  {Whitlock}}]{Wintermantel2020Cellular}%
  \BibitemOpen
  \bibfield  {author} {\bibinfo {author} {\bibfnamefont {T.~M.}\ \bibnamefont
  {Wintermantel}}, \bibinfo {author} {\bibfnamefont {Y.}~\bibnamefont {Wang}},
  \bibinfo {author} {\bibfnamefont {G.}~\bibnamefont {Lochead}}, \bibinfo
  {author} {\bibfnamefont {S.}~\bibnamefont {Shevate}}, \bibinfo {author}
  {\bibfnamefont {G.~K.}\ \bibnamefont {Brennen}},\ and\ \bibinfo {author}
  {\bibfnamefont {S.}~\bibnamefont {Whitlock}},\ }\bibfield  {title} {\bibinfo
  {title} {Unitary and nonunitary quantum cellular automata with {Rydberg}
  arrays},\ }\href {https://doi.org/10.1103/PhysRevLett.124.070503} {\bibfield
  {journal} {\bibinfo  {journal} {Phys. Rev. Lett.}\ }\textbf {\bibinfo
  {volume} {124}},\ \bibinfo {pages} {070503} (\bibinfo {year}
  {2020})}\BibitemShut {NoStop}%
\bibitem [{\citenamefont {Klocke}\ \emph {et~al.}(2021)\citenamefont {Klocke},
  \citenamefont {Wintermantel}, \citenamefont {Lochead}, \citenamefont
  {Whitlock},\ and\ \citenamefont {Buchhold}}]{klocke2021hydrodynamic}%
  \BibitemOpen
  \bibfield  {author} {\bibinfo {author} {\bibfnamefont {K.}~\bibnamefont
  {Klocke}}, \bibinfo {author} {\bibfnamefont {T.}~\bibnamefont
  {Wintermantel}}, \bibinfo {author} {\bibfnamefont {G.}~\bibnamefont
  {Lochead}}, \bibinfo {author} {\bibfnamefont {S.}~\bibnamefont {Whitlock}},\
  and\ \bibinfo {author} {\bibfnamefont {M.}~\bibnamefont {Buchhold}},\
  }\bibfield  {title} {\bibinfo {title} {Hydrodynamic stabilization of
  self-organized criticality in a driven {Rydberg} gas},\ }\href
  {https://journals.aps.org/prl/abstract/10.1103/PhysRevLett.126.123401}
  {\bibfield  {journal} {\bibinfo  {journal} {Phys. Rev. Lett.}\ }\textbf
  {\bibinfo {volume} {126}},\ \bibinfo {pages} {123401} (\bibinfo {year}
  {2021})}\BibitemShut {NoStop}%
\bibitem [{\citenamefont {Gambetta}\ \emph {et~al.}(2019)\citenamefont
  {Gambetta}, \citenamefont {Carollo}, \citenamefont {Marcuzzi}, \citenamefont
  {Garrahan},\ and\ \citenamefont {Lesanovsky}}]{gambetta2019}%
  \BibitemOpen
  \bibfield  {author} {\bibinfo {author} {\bibfnamefont {F.~M.}\ \bibnamefont
  {Gambetta}}, \bibinfo {author} {\bibfnamefont {F.}~\bibnamefont {Carollo}},
  \bibinfo {author} {\bibfnamefont {M.}~\bibnamefont {Marcuzzi}}, \bibinfo
  {author} {\bibfnamefont {J.~P.}\ \bibnamefont {Garrahan}},\ and\ \bibinfo
  {author} {\bibfnamefont {I.}~\bibnamefont {Lesanovsky}},\ }\bibfield  {title}
  {\bibinfo {title} {Discrete time crystals in the absence of manifest
  symmetries or disorder in open quantum systems},\ }\href
  {https://doi.org/10.1103/PhysRevLett.122.015701} {\bibfield  {journal}
  {\bibinfo  {journal} {Phys. Rev. Lett.}\ }\textbf {\bibinfo {volume} {122}},\
  \bibinfo {pages} {015701} (\bibinfo {year} {2019})}\BibitemShut {NoStop}%
\bibitem [{\citenamefont {Wadenpfuhl}\ and\ \citenamefont
  {Adams}(2023)}]{Wadenpfuhl2023Synchronization}%
  \BibitemOpen
  \bibfield  {author} {\bibinfo {author} {\bibfnamefont {K.}~\bibnamefont
  {Wadenpfuhl}}\ and\ \bibinfo {author} {\bibfnamefont {C.~S.}\ \bibnamefont
  {Adams}},\ }\bibfield  {title} {\bibinfo {title} {Emergence of
  synchronization in a driven-dissipative hot {Rydberg} vapor},\ }\href
  {https://doi.org/10.1103/PhysRevLett.131.143002} {\bibfield  {journal}
  {\bibinfo  {journal} {Phys. Rev. Lett.}\ }\textbf {\bibinfo {volume} {131}},\
  \bibinfo {pages} {143002} (\bibinfo {year} {2023})}\BibitemShut {NoStop}%
\bibitem [{\citenamefont {Ding}\ \emph {et~al.}(2024)\citenamefont {Ding},
  \citenamefont {Bai}, \citenamefont {Liu}, \citenamefont {Shi}, \citenamefont
  {Guo}, \citenamefont {Li},\ and\ \citenamefont {Adams}}]{ding2024ergodicity}%
  \BibitemOpen
  \bibfield  {author} {\bibinfo {author} {\bibfnamefont {D.}~\bibnamefont
  {Ding}}, \bibinfo {author} {\bibfnamefont {Z.}~\bibnamefont {Bai}}, \bibinfo
  {author} {\bibfnamefont {Z.}~\bibnamefont {Liu}}, \bibinfo {author}
  {\bibfnamefont {B.}~\bibnamefont {Shi}}, \bibinfo {author} {\bibfnamefont
  {G.}~\bibnamefont {Guo}}, \bibinfo {author} {\bibfnamefont {W.}~\bibnamefont
  {Li}},\ and\ \bibinfo {author} {\bibfnamefont {C.~S.}\ \bibnamefont
  {Adams}},\ }\bibfield  {title} {\bibinfo {title} {Ergodicity breaking from
  {Rydberg} clusters in a driven-dissipative many-body system},\ }\href
  {https://www.science.org/doi/full/10.1126/sciadv.adl5893} {\bibfield
  {journal} {\bibinfo  {journal} {Science advances}\ }\textbf {\bibinfo
  {volume} {10}},\ \bibinfo {pages} {eadl5893} (\bibinfo {year}
  {2024})}\BibitemShut {NoStop}%
\bibitem [{\citenamefont {Wu}\ \emph {et~al.}(2023{\natexlab{a}})\citenamefont
  {Wu}, \citenamefont {Wang}, \citenamefont {Fan~Yang}, \citenamefont {Liang},
  \citenamefont {Tey}, \citenamefont {Li}, \citenamefont {Pohl},\ and\
  \citenamefont {You}}]{wu2023}%
  \BibitemOpen
  \bibfield  {author} {\bibinfo {author} {\bibfnamefont {X.}~\bibnamefont
  {Wu}}, \bibinfo {author} {\bibfnamefont {Z.}~\bibnamefont {Wang}}, \bibinfo
  {author} {\bibfnamefont {R.~G.}\ \bibnamefont {Fan~Yang}}, \bibinfo {author}
  {\bibfnamefont {C.}~\bibnamefont {Liang}}, \bibinfo {author} {\bibfnamefont
  {M.~K.}\ \bibnamefont {Tey}}, \bibinfo {author} {\bibfnamefont
  {X.}~\bibnamefont {Li}}, \bibinfo {author} {\bibfnamefont {T.}~\bibnamefont
  {Pohl}},\ and\ \bibinfo {author} {\bibfnamefont {L.}~\bibnamefont {You}},\
  }\bibfield  {title} {\bibinfo {title} {Observation of a dissipative time
  crystal in a strongly interacting {Rydberg} gas},\ }\href
  {https://arxiv.org/abs/2305.20070} {\bibfield  {journal} {\bibinfo  {journal}
  {arXiv:2305.20070}\ } (\bibinfo {year} {2023}{\natexlab{a}})}\BibitemShut
  {NoStop}%
\bibitem [{\citenamefont {Liu}\ \emph {et~al.}(2024{\natexlab{a}})\citenamefont
  {Liu}, \citenamefont {Zhang}, \citenamefont {Liu}, \citenamefont {Zhang},
  \citenamefont {Zhang}, \citenamefont {Shao}, \citenamefont {Li},
  \citenamefont {Chen}, \citenamefont {Ma}, \citenamefont {Han} \emph
  {et~al.}}]{liu2024higher}%
  \BibitemOpen
  \bibfield  {author} {\bibinfo {author} {\bibfnamefont {B.}~\bibnamefont
  {Liu}}, \bibinfo {author} {\bibfnamefont {L.-H.}\ \bibnamefont {Zhang}},
  \bibinfo {author} {\bibfnamefont {Z.-K.}\ \bibnamefont {Liu}}, \bibinfo
  {author} {\bibfnamefont {J.}~\bibnamefont {Zhang}}, \bibinfo {author}
  {\bibfnamefont {Z.-Y.}\ \bibnamefont {Zhang}}, \bibinfo {author}
  {\bibfnamefont {S.-Y.}\ \bibnamefont {Shao}}, \bibinfo {author}
  {\bibfnamefont {Q.}~\bibnamefont {Li}}, \bibinfo {author} {\bibfnamefont
  {H.-C.}\ \bibnamefont {Chen}}, \bibinfo {author} {\bibfnamefont
  {Y.}~\bibnamefont {Ma}}, \bibinfo {author} {\bibfnamefont {T.-Y.}\
  \bibnamefont {Han}}, \emph {et~al.},\ }\bibfield  {title} {\bibinfo {title}
  {Higher-order and fractional discrete time crystals in floquet-driven
  {Rydberg} atoms},\ }\href {https://doi.org/10.48550/arXiv.2402.13657}
  {\bibfield  {journal} {\bibinfo  {journal} {arXiv preprint arXiv:2402.13657}\
  } (\bibinfo {year} {2024}{\natexlab{a}})}\BibitemShut {NoStop}%
\bibitem [{\citenamefont {Liu}\ \emph {et~al.}(2024{\natexlab{b}})\citenamefont
  {Liu}, \citenamefont {Zhang}, \citenamefont {Liu}, \citenamefont {Zhang},
  \citenamefont {Zhang}, \citenamefont {Shao}, \citenamefont {Li},
  \citenamefont {Chen}, \citenamefont {Ma}, \citenamefont {Han} \emph
  {et~al.}}]{liu2024bifurcation}%
  \BibitemOpen
  \bibfield  {author} {\bibinfo {author} {\bibfnamefont {B.}~\bibnamefont
  {Liu}}, \bibinfo {author} {\bibfnamefont {L.-H.}\ \bibnamefont {Zhang}},
  \bibinfo {author} {\bibfnamefont {Z.-K.}\ \bibnamefont {Liu}}, \bibinfo
  {author} {\bibfnamefont {J.}~\bibnamefont {Zhang}}, \bibinfo {author}
  {\bibfnamefont {Z.-Y.}\ \bibnamefont {Zhang}}, \bibinfo {author}
  {\bibfnamefont {S.-Y.}\ \bibnamefont {Shao}}, \bibinfo {author}
  {\bibfnamefont {Q.}~\bibnamefont {Li}}, \bibinfo {author} {\bibfnamefont
  {H.-C.}\ \bibnamefont {Chen}}, \bibinfo {author} {\bibfnamefont
  {Y.}~\bibnamefont {Ma}}, \bibinfo {author} {\bibfnamefont {T.-Y.}\
  \bibnamefont {Han}}, \emph {et~al.},\ }\bibfield  {title} {\bibinfo {title}
  {Bifurcation of time crystals in driven and dissipative {Rydberg} atomic
  gas},\ }\href {https://doi.org/10.48550/arXiv.2402.13644} {\bibfield
  {journal} {\bibinfo  {journal} {arXiv preprint arXiv:2402.13644}\ } (\bibinfo
  {year} {2024}{\natexlab{b}})}\BibitemShut {NoStop}%
\bibitem [{\citenamefont {Wu}\ \emph {et~al.}(2023{\natexlab{b}})\citenamefont
  {Wu}, \citenamefont {Wang}, \citenamefont {Yang}, \citenamefont {Gao},
  \citenamefont {Liang}, \citenamefont {Tey}, \citenamefont {Li}, \citenamefont
  {Pohl},\ and\ \citenamefont {You}}]{wu2023observation}%
  \BibitemOpen
  \bibfield  {author} {\bibinfo {author} {\bibfnamefont {X.}~\bibnamefont
  {Wu}}, \bibinfo {author} {\bibfnamefont {Z.}~\bibnamefont {Wang}}, \bibinfo
  {author} {\bibfnamefont {F.}~\bibnamefont {Yang}}, \bibinfo {author}
  {\bibfnamefont {R.}~\bibnamefont {Gao}}, \bibinfo {author} {\bibfnamefont
  {C.}~\bibnamefont {Liang}}, \bibinfo {author} {\bibfnamefont {M.~K.}\
  \bibnamefont {Tey}}, \bibinfo {author} {\bibfnamefont {X.}~\bibnamefont
  {Li}}, \bibinfo {author} {\bibfnamefont {T.}~\bibnamefont {Pohl}},\ and\
  \bibinfo {author} {\bibfnamefont {L.}~\bibnamefont {You}},\ }\bibfield
  {title} {\bibinfo {title} {Observation of a dissipative time crystal in a
  strongly interacting {Rydberg} gas},\ }\href
  {https://arxiv.org/abs/2305.20070} {\bibfield  {journal} {\bibinfo  {journal}
  {arXiv preprint arXiv:2305.20070}\ } (\bibinfo {year}
  {2023}{\natexlab{b}})}\BibitemShut {NoStop}%
\bibitem [{\citenamefont {Zhang}\ \emph {et~al.}(2022)\citenamefont {Zhang},
  \citenamefont {Liu}, \citenamefont {Liu}, \citenamefont {Zhang},
  \citenamefont {Guo}, \citenamefont {Ding},\ and\ \citenamefont
  {Shi}}]{zhangRydberg}%
  \BibitemOpen
  \bibfield  {author} {\bibinfo {author} {\bibfnamefont {L.-H.}\ \bibnamefont
  {Zhang}}, \bibinfo {author} {\bibfnamefont {Z.-K.}\ \bibnamefont {Liu}},
  \bibinfo {author} {\bibfnamefont {B.}~\bibnamefont {Liu}}, \bibinfo {author}
  {\bibfnamefont {Z.-Y.}\ \bibnamefont {Zhang}}, \bibinfo {author}
  {\bibfnamefont {G.-C.}\ \bibnamefont {Guo}}, \bibinfo {author} {\bibfnamefont
  {D.-S.}\ \bibnamefont {Ding}},\ and\ \bibinfo {author} {\bibfnamefont
  {B.-S.}\ \bibnamefont {Shi}},\ }\bibfield  {title} {\bibinfo {title} {Rydberg
  microwave-frequency-comb spectrometer},\ }\href
  {https://doi.org/10.1103/PhysRevApplied.18.014033} {\bibfield  {journal}
  {\bibinfo  {journal} {Phys. Rev. Applied}\ }\textbf {\bibinfo {volume}
  {18}},\ \bibinfo {pages} {014033} (\bibinfo {year} {2022})}\BibitemShut
  {NoStop}%
\bibitem [{\citenamefont {Liu}\ \emph {et~al.}(2022)\citenamefont {Liu},
  \citenamefont {Zhang}, \citenamefont {Liu}, \citenamefont {Zhang},
  \citenamefont {Zhu}, \citenamefont {Gao}, \citenamefont {Guo}, \citenamefont
  {Ding},\ and\ \citenamefont {Shi}}]{liuHighly}%
  \BibitemOpen
  \bibfield  {author} {\bibinfo {author} {\bibfnamefont {B.}~\bibnamefont
  {Liu}}, \bibinfo {author} {\bibfnamefont {L.-H.}\ \bibnamefont {Zhang}},
  \bibinfo {author} {\bibfnamefont {Z.-K.}\ \bibnamefont {Liu}}, \bibinfo
  {author} {\bibfnamefont {Z.-Y.}\ \bibnamefont {Zhang}}, \bibinfo {author}
  {\bibfnamefont {Z.-H.}\ \bibnamefont {Zhu}}, \bibinfo {author} {\bibfnamefont
  {W.}~\bibnamefont {Gao}}, \bibinfo {author} {\bibfnamefont {G.-C.}\
  \bibnamefont {Guo}}, \bibinfo {author} {\bibfnamefont {D.-S.}\ \bibnamefont
  {Ding}},\ and\ \bibinfo {author} {\bibfnamefont {B.-S.}\ \bibnamefont
  {Shi}},\ }\bibfield  {title} {\bibinfo {title} {Highly sensitive measurement
  of a megahertz rf electric field with a {Rydberg}-atom sensor},\ }\href
  {https://doi.org/10.1103/PhysRevApplied.18.014045} {\bibfield  {journal}
  {\bibinfo  {journal} {Phys. Rev. Applied}\ }\textbf {\bibinfo {volume}
  {18}},\ \bibinfo {pages} {014045} (\bibinfo {year} {2022})}\BibitemShut
  {NoStop}%
\end{thebibliography}%

\section*{Methods}
In the methods section, the experimental setup is introduced in more details. Then, we show the Hamiltonian of system and drive the Landau-type effective potential to obtain the conditions of U(1) symmetry breaking. Finally, we introduce massless and mass excitations, and Lorentz-invariant effective action.

\subsection*{Experimental setup}
In the experiments, we employed a three-photon EIT scheme to excite the Rydberg atoms, as illustrated in Fig.~\ref{setup}(b). The 852 nm external-cavity diode laser (ECDL) was frequency-locked using the saturated absorption spectrum as a reference signal. Another ECDL at the wavelength of 1470 nm was locked by referencing the two-photon spectrum. The 780 nm ECDL, acting as the coupling laser, was amplified using a tapered amplifier with an output power of 2 W. The probe laser was split into two beams, referred to as probe 1 and probe 2. The probe laser beams passed through a 7-cm vapor cell, with the dressing and coupling beams counter-propagating with respect to the probe beams. The probe beams were focused into the cell, with a $1/e^2$ waist radius of approximately 200 $\mu$m and a probe intensity of about 70 $\mu$W. The dressing and coupling beams were focused with a $1/e^2$ waist radius of around 500 $\mu$m, with powers of 16.8 mW and 1.5 W, respectively. 

A circular copper plate, 3 mm thick and with a diameter of 120 mm, was used in the experiment. The separation between the plates was set to 40 mm, with the Cesium cell placed at the center between them. An arbitrary function generator (AFG) was used to generate the RF field applied to the two electrodes, with the RF frequency set at $\omega_\text{RF}$ = $2\pi \times$7.1 MHz. The probe transmissions were measured by an oscilloscope after the signals passed through photoelectric detectors. To scan external parameters and obtain the phase diagram of the system's response, the coupling laser, oscilloscope, and AFG were all connected to a computer for synchronized control.

\subsection*{Hamiltonian}
The Hamiltonian of the system, incorporating both light and atoms, is expressed as 
\begin{equation}
\begin{aligned}
H &= \hbar \Delta_{1} \hat{a}_{p_1}^\dagger \hat{a}_{p_1} + \hbar \Delta_{2} \hat{a}_{p_2}^\dagger \hat{a}_{p_2} + \hbar \delta_1 \sum_i \hat{\sigma}_{R_1 R_1}^{(i)} \\
& + \hbar \delta_2 \sum_i \hat{\sigma}_{R_2 R_2}^{(i)} + \hbar g_{p_1} \left( \hat{a}_{p_1}^\dagger \sum_i \hat{\sigma}_{g R_1}^{(i)} + \hat{a}_{p_1} \sum_i \hat{\sigma}_{R_1 g}^{(i)} \right) 
\\& + \hbar g_{p_2} \left( \hat{a}_{p_2}^\dagger \sum_i \hat{\sigma}_{g R_2}^{(i)} + \hat{a}_{p_2} \sum_i \hat{\sigma}_{R_2 g}^{(i)} \right) \\
&+ \sum_{i\neq j} V_{ij} \hat{\sigma}_{R_1 R_1}^{(i)} \hat{\sigma}_{R_1 R_1}^{(j)}
+ \sum_{i\neq j} V_{ij} \hat{\sigma}_{R_2 R_2}^{(i)} \hat{\sigma}_{R_2 R_2}^{(j)}
\\& +2 \sum_{i\neq j} V_{ij} \hat{\sigma}_{R_1 R_1}^{(i)} \hat{\sigma}_{R_2 R_2}^{(j)}.
\end{aligned}
\end{equation}
where $\hat{a}_{p1(p2)}^\dagger$ and $\hat{a}_{p1(p2)}$ are the creation and annihilation operators of probe 1 (2) field, $\hat{\sigma}_{m n} = \left| m \right\rangle {{\left\langle n \right|}}$ ($m$, $n$ = $g$, $R_1$, $R_2$), $g_{p1(p2)}$ represents the coupling strength between probe 1(2) and the atomic transition between the ground state $\ket{g}$ and the Rydberg state $\ket{R_{1(2)}}$. The energy shifts due to the applied RF electric field are given by $\delta_1 = -\frac{1}{4}a_1 E_{\rm{RF}}^2$ and $\delta_2 = -\frac{1}{4}a_2 E_{\rm{RF}}^2$, where $a_1$ and $a_2$ are the polarizabilities of the respective Rydberg states, and $E_{\rm{RF}}$ is the amplitude of the RF electric field. This Hamiltonian describes the interaction between two light fields and an ensemble of atoms, where the atoms can be excited from a common ground state $\ket{g}$ to two different Rydberg states $\ket{R_1}$ and $\ket{R_2}$. The two light fields, with respective frequencies $\omega_{p1}$ and $\omega_{p2}$, drive transitions between the ground state and the RF field-shifted Rydberg states with detunings $\Delta_1$ and $\Delta_2$, respectively. 

\begin{figure*}
\centering
\includegraphics[width=2.08\columnwidth]{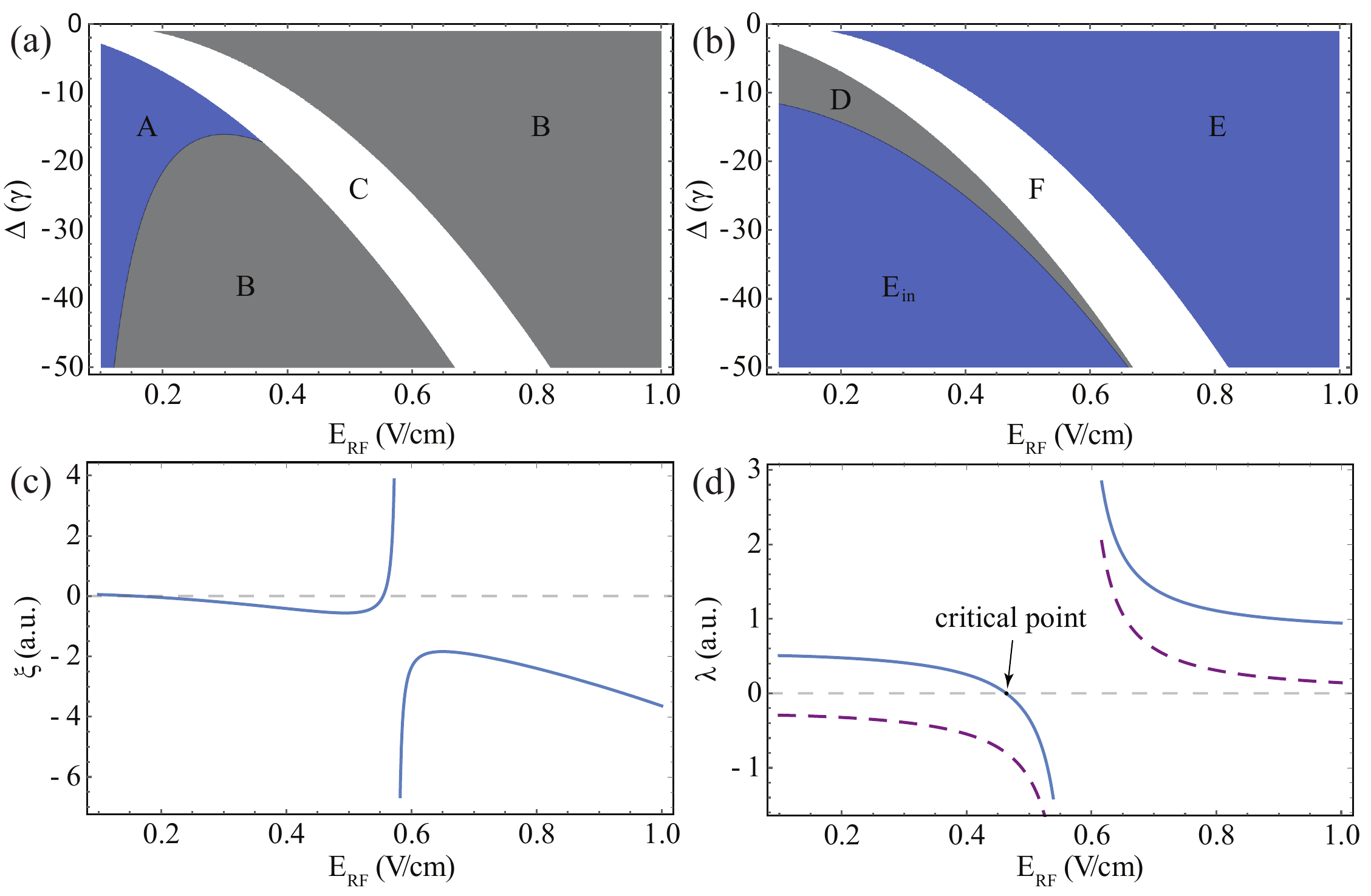}\\
\caption{\textbf{Contour plot of $\xi$ and $\lambda$.} Map of $\xi$ (a) and $\lambda$ (b) versus $\Delta$ and $E_{\text{RF}}$. The blue and gray colors represent $\xi, \lambda>0$ [A regime in (a) and E regime in (b)] and $\xi, \lambda<0$ [B regime in (a) and D regime in (b)]. The solid lines in (a) and (b) are the boundary that represent the case of $\xi, \lambda=0$. The areas of C and F are conventional symmetry regions where the Higgs and Goldstone modes are absent. In these simulations, we set $N$= 4, $g$ = 4$\gamma$, $\hbar$ = 1, and $V$ = 2$\gamma$. (c) and (d) correspond to the respective cases under $\Delta = -30\gamma$. In (d), the marked point is the critical point of U(1)-symmetry breaking. The purple dashed line corresponds to the case of $V$ = 0.}
\label{U(1) symmetry}
\end{figure*}

\subsection*{Landau-type effective potential}
To derive the effective potential, we first insert a mean-field ansatz for the probe fields $\langle \hat{a}_{p_1} \rangle = \alpha_1$, $\quad \langle \hat{a}_{p_2} \rangle = \alpha_2$, and for the atomic state populations $\langle \hat{\sigma}_{R_1 R_1} \rangle = $$\psi_1^2$,  $\langle \hat{\sigma}_{R_2 R_2} \rangle = \psi_2^2$,  $\langle \hat{\sigma}_{g g} \rangle = \psi_0^2$, where $\alpha_1$, $\alpha_2$, and $\psi_1$, $\psi_2$, $\psi_0$ are real-valued order parameters that describe the amplitudes of the probe fields and atomic states. Due to the particle conservation, the condition $\psi_0^2+\psi_1^2+\psi_2^2=1$ is satisfied. For U(1)–symmetric case, we set $\Delta_1 = \Delta_2 = \Delta-\delta$, $\delta_1 = \delta_2 = \delta$ and $g_{p_1}=g_{p_2} =g$. By substituting these mean-field expressions into the Hamiltonian, we obtain an effective potential that depends on the atomic and photonic order parameters. This effective potential becomes as: 
\begin{equation}
\begin{aligned}
V_{\text{eff}}(\alpha_1, \alpha_2, \psi_1, \psi_2) & = \hbar \Delta (\alpha_1^2 + \alpha_2^2) + \hbar \delta N (\psi_1^2 + \psi_2^2) \\ & + 2\hbar g N (\alpha_1 \psi_1 + \alpha_2 \psi_2) \psi_0 \\ & + N(N-1)V(\psi_1^2 + \psi_2^2)^2.
\end{aligned}
\end{equation}
Given that the light field reaches its steady state much faster than the atomic fields (i.e., $\partial V_{\text{eff}}/{\partial \alpha_i} = 0
$), we adiabatically eliminate the photon fields $\alpha_1$ and $\alpha_2$. This leads to the following condition $\alpha_1 = -g N \psi_1 \psi_0/{\Delta}$,  $\alpha_2 = -g N \psi_2 \psi_0/{\Delta}$. Substituting these expressions back into the effective potential gives: 
\begin{equation}
\begin{aligned}
    V_{\text{eff}}(\psi_1, \psi_2) &= \hbar \delta N (\psi_1^2 + \psi_2^2) - \frac{ \hbar g^2 N^2}{\Delta}  \psi_0^2 \left( \psi_1^2 + \psi_2^2 \right) \\ & +N(N-1)V(\psi_1^2 + \psi_2^2)^2.
\end{aligned}
\end{equation}
We define a combined order parameter $\psi=\psi_1 + i\psi_2$, the potential can be rewritten in the standard form of a Landau-type effective potential: 
\begin{equation}
V_{\text{eff}}(\psi) = \xi |\psi|^2 + \lambda |\psi|^4
\end{equation}
where $\xi =\hbar\delta N- \hbar g^2 N^2/(\Delta-\delta)$ and $\lambda=\hbar g^2 N^2/(\Delta-\delta)+N(N-1)V$ are coefficients. Figures.~\ref{U(1) symmetry}(a-b) show the contour plots of $\xi$ and $\lambda$ versus $\Delta$ and $E_{\text{RF}}$. In the regime where $\xi<$ 0 [B regime in Fig.~\ref{U(1) symmetry} (a)] and $\lambda > 0$ [E regime in Fig.~\ref{U(1) symmetry} (b)], the effective potential $V_{\text{eff}}(\psi)$ takes on the shape of a Mexican hat, leading to the U(1)-symmetry breaking. In contrast, when these conditions are not met, the U(1) symmetry remains unbroken. Figures.~\ref{U(1) symmetry}(c-d) display results under the condition $\Delta = -30\gamma$. In Fig.~\ref{U(1) symmetry}(d), the intersection point with the axis corresponds to the critical point $E_{\text{RF,c}}$. For $E_{\text{RF}}<E_{\text{RF,c}}$, $\lambda>0$, resulting in a Mexican hat-shaped potential. Conversely, when $E_{\text{RF}}>E_{\text{RF,c}}$, $\lambda<0$, leading to a conventional potential shape. The non-zero interaction strength $V$ makes $\text{E}_\text{in}$ regime in Fig.~\ref{U(1) symmetry} (b) in symmetry-broken, otherwise this regime is in symmetry [$\lambda<0$, see also the purple dashed line in Fig.~\ref{U(1) symmetry} (d)]. These results conclude that the interaction between Rydberg atoms induces additional Higgs and Goldstone modes in $\text{E}_\text{in}$ regime.

\subsection*{Massless and mass excitations}
The Lagrangian for a complex scalar field $\psi=\psi_1 + i\psi_2$ takes the form: 
\begin{equation}
L = (\partial_\mu \psi)^* (\partial^\mu \psi) - \xi \psi^* \psi - \lambda (\psi^* \psi)^2
\end{equation}
This Lagrangian has a U(1) symmetry,  where the field $\psi$ transforms under a phase shift as $\psi \rightarrow \psi e^{i\alpha}$. $\xi<0$ leads to spontaneous symmetry breaking as the potential no longer has its minimum at $\psi=0$, where $\partial V_{\text{eff}}/{\partial \psi} = 0$ gives that the field at the potential minima satisfies $\psi^2=-\xi/\lambda = \nu^2$. To avoid having non-zero fields in the vacuum, we transform the complex scalar field $\psi=\psi_1 + i\psi_2$ to a shifted field $\phi=\psi-\nu=\phi_1 + i\phi_2$. By considering the shifted field, the Lagrangian is transformed into:
\begin{equation}
\mathcal{L} =  (\partial_\mu \phi_1)^2 + (\partial_\mu \phi_2)^2  + 2\xi \phi_2^2
\end{equation}
thus, we obtain the effective mass for fields $\phi_1$ and $\phi_2$ as:
\begin{equation}
\begin{aligned}
 & m_{\text{Higgs}} =\sqrt{4\hbar N( g^2N/(\Delta-\delta)-\delta)} \\
 & m_{\text{Goldstone}} =0
\end{aligned}
\end{equation}

\subsection*{Lorentz-invariant effective action}
To capture the time-dependent fluctuations about the state $\psi$, we need to add dynamic terms to the effective potential. The time-dependent Landau-Ginzburg effective action density is expressed as \cite{leonard2017monitoring}:
\begin{equation}
\begin{aligned}
S_{\text{dyn}} &= K_1 \left( \psi^*(r,t) \frac{\partial}{\partial t} \psi(r,t) 
- \psi(r,t) \frac{\partial}{\partial t} \psi^*(r,t) \right) \\
&+ K_2 \left( \frac{\partial}{\partial t} \psi^*(r,t) \right) \left( \frac{\partial}{\partial t} \psi(r,t) \right)
\end{aligned}
\end{equation}
The first term describes the phase evolution of the field, associated with the coefficient $K_1$. The second term with a coefficient $K_2$ is related to the square of the time derivative of the field and represents the dissipation or damping. By substituting $\psi=\psi_1 + i\psi_2$, we obtain 
\begin{equation}
S_{\text{dyn}} =2iK_1 (\psi_1 \partial_t \psi_2 - \psi_2 \partial_t \psi_1) 
+ K_2 \left[ (\partial_t \psi_1)^2 + (\partial_t \psi_2)^2 \right]
\end{equation}
In our system, the Hamiltonian is invariant under the exchange of the two probes, $\alpha_1$ and $\alpha_2$, as well as the corresponding Rydberg collective excitation modes $\psi_1 $ and $\psi_2$. This leads to the presence of particle-hole symmetry in the system, which requires $K_1$ = 0 \cite{pekker2015amplitude}. With the fully particle-hole symmetric condition [$K_1$ = 0, $K_2\neq 0$], we get two orthogonal and distinct modes for amplitude and phase fluctuations.

\section*{Acknowledgements}
We acknowledge funding from the National Key R and D Program of China (Grant No. 2022YFA1404002), the National Natural Science Foundation of China (Grant Nos. U20A20218, 61525504, and 61435011), the Anhui Initiative in Quantum Information Technologies (Grant No. AHY020200), and the Major Science and Technology Projects in Anhui Province (Grant No. 202203a13010001).

\section*{Author contributions statement}
D.-S.D. conceived the idea for the study. B.L. conducted the physical experiments. D.-S.D and B.L. developed the theoretical model and written the manuscript. The research was supervised by D.-S.D. All authors contributed to discussions regarding the results and the analysis contained in the manuscript.

\section*{Competing interests}
The authors declare no competing interests.

\end{document}